\documentclass[12pt]{article}
\pdfoutput=1
\usepackage{latexsym, graphicx} 
\usepackage{amsmath}
\numberwithin{equation}{section}
\usepackage{amssymb}
\usepackage{amsfonts}
\usepackage{dsfont}
\usepackage{caption}
\usepackage{subcaption}
\usepackage{tensor}
\usepackage{cite}
\usepackage[utf8]{inputenc}
\usepackage{tikz}
\usetikzlibrary{decorations.pathmorphing}
\tikzset{snake it/.style={decorate, decoration=snake}}
\usepackage{IEEEtrantools}

\definecolor{asparagus}{rgb}{0.53, 0.85, 0.30}
\definecolor{alizarin}{rgb}{0.9, 0.1, 0.1}
\definecolor{azure}{rgb}{0, 0.1, 1.0}
\usepackage[colorlinks=true, linkcolor=azure, citecolor=asparagus, urlcolor=alizarin, bookmarks=true]{hyperref}
\newcommand{\arXiv}[1]{\href{http://www.arXiv.org/abs/#1}{#1}}

\makeatletter
\renewcommand\section{\@startsection {section}{1}{\z@}%
	{-3.5ex \@plus -1ex \@minus -.2ex}
	{2.3ex \@plus.2ex}%
	{\normalfont\large\bfseries}}
\renewcommand\subsection{\@startsection{subsection}{2}{\z@}%
	{-3.25ex\@plus -1ex \@minus -.2ex}%
	{1.5ex \@plus .2ex}%
	{\normalfont\bfseries}}
\makeatother

\begin{document}	

\begin{titlepage}
	\vfill
	\begin{center}
		{\LARGE \bf De Sitter-invariant States from Holography}
		
		\vskip 20mm
		
		{\large K\'evin Nguyen}
		
		\vskip 15mm
		
		Theoretische Natuurkunde, Vrije Universiteit Brussel (VUB), and\\
		International Solvay Institutes, Pleinlaan 2, B-1050 Brussels, Belgium \\
		
		\vskip 15mm
		
		{\small\noindent  {\tt Kevin.Huy.D.Nguyen@vub.be}}
		
	\end{center}
	\vfill

	\begin{center}
		{\bf ABSTRACT}
		\vspace{3mm}
	\end{center}
	
	 A class of invariant states under de Sitter isometries is constructed in $d$-dimensional Conformal Field Theories from the universal sector of AdS/CFT dualities. These states extend the Mottola-Allen $\alpha$-vacua to theories containing scalar primary operators of arbitrary scaling dimensions, and are proven to explicitly break conformal symmetry. Two-point correlators in these states are shown to satisfy a generalized spectral decomposition in terms of free massive propagators. It is also pointed out that the spectral decomposition of generic Euclidean correlators is known in the mathematics literature as Olevsky index transform, of which computational use can be made. 
	
	\vfill
	
\end{titlepage}	

\tableofcontents

\section{Introduction and summary}
Quantum Field Theory (QFT) in an expanding spacetime is of central importance in modern theoretical cosmology, as it directly enters the physics of the early Universe. Thanks to the inflationary phase that the latter underwent, imprints of quantum phenomena having taken place in this early period are still measured today in the Cosmic Microwave Background \cite{Parker,Baumann:2009ds}.\\

Much effort has been invested into the exploration of QFT in \textit{de Sitter space} (dS). It is a maximally symmetric curved spacetime that possesses as many isometries as Minkowski space. It can also be used to describe an exponentially expanding FLRW universe with flat spatial sections ($k=0$) and positive cosmological constant, such that it constitutes a very natural cosmological toy model in which quantum field theory can be investigated. Although this framework provides computational simplifications, one is still facing issues of infrared divergences and secular terms coming from higher-loop corrections, in the perturbative approach to interacting theories. See \cite{Polyakov:2012uc,Akhmedov:2013vka} and references therein. The story is however different if one considers theories with extra symmetries such as conformal symmetry (CFT) that allows non-perturbative results to be extracted. Obviously one can already infer much information about the \textit{conformal vacuum} from non-perturbative flat space results as de Sitter space is conformally flat. As an example, the functional form of the renormalized stress-energy tensor can be computed exactly and reliable exploration of the backreaction problem may be performed. A series expansion away from conformality then provides a good starting point to study backreaction induced by more realistic interacting QFTs. See \cite{PerezNadal:2008ju} and references therein. In the spirit of exploring non-perturbative physics of de Sitter CFTs, techniques relying on \textit{holography}\footnote{Holography, also called gauge/gravity duality or AdS/CFT, is in its weaker form a well-tested duality between classical field theories in spacetimes with asymptotic anti-de-Sitter geometry (AdS) and strongly interacting conformal field theories living on the AdS boundary whose intrinsic geometry is kept fixed. See \cite{Ammon:2015wua} for a modern review. In the case of interest to us, the fixed boundary geometry where the CFT lives is de Sitter space.} started being applied in recent years following the early work of Hawking, Maldacena and Strominger \cite{Hawking:2000da}. A sample of holographic studies, such as Schwinger pair production in de Sitter CFTs, includes \cite{Koyama:2001rf,Hirayama:2006jn,Hutasoit:2009sc,Marolf:2010tg,Andrade:2011nh,Fischler:2013fba,Fischler:2014ama,Fischler:2014tka,Chu:2016uwi}.\\ 

Already in the late sixties Chernikov and Tagirov discovered that free massive QFTs in de Sitter space admit a whole family of de Sitter-invariant vacua \cite{Chernikov:1968zm} which were subsequently studied by Mottola and Allen \cite{Mottola:1984ar,Allen:1985ux}. We will refer to these states as Mottola-Allen (MA) vacua or $\alpha$-\textit{vacua}. For a nice group theoretic construction of these non-trivial vacua, see \cite{Joung:2006gj}. Whether $\alpha$-vacua are physically relevant is still the subject of debates. It is often argued that for cosmological purposes the most natural one is the \textit{Euclidean/Bunch-Davies vacuum} (or \textit{conformal vacuum} in the case of the conformally coupled scalar) in which positive frequencies are defined with respect to a free falling observer at early conformal time (see Appendix \ref{Appendix:freeQFT}). Futhermore the Euclidean vacuum is the unique \textit{Hadamard state} among the whole family, which makes it the rightful vacuum of the theory according to algebraic approaches to QFT \cite{Kay:1988mu,Hollands:2014eia}. However, in the context of non-equilibrium QFTs, $\alpha$-vacua are argued to encode high-energy/irrelevant corrections to the initial state of the inflationary Universe \cite{Einhorn:2003xb,deBoer:2004nd,Schalm:2004xg,Collins:2005cm,Polyakov:2012uc}. They also play a central role in the development of the dS/CFT duality conjectured to describe quantum gravity in de Sitter space from a dual conformal field theory living on the asymptotic null boundaries $\mathcal{I^+}/\mathcal{I^-}$ \cite{Bousso:2001mw,Spradlin:2001nb}.\\ 

In this paper we extend the Mottola-Allen construction of de Sitter-invariant states to holographic CFTs having primary operators of arbitrary scaling dimensions above the unitarity bound, $\Delta \geq d/2-1$.\footnote{The conformally coupled scalar is a particular case of scaling dimension $\Delta=\frac{d}{2}-1$ that actually saturates this bound.} These new states generically differ from the conformal vacuum and will be shown to \textit{explicitly break conformal symmetry}.\footnote{One may wonder which states of the corresponding flat space CFT these are mapped to under conformal transformation from de Sitter to Minkowski space. Although we do not have an explicit answer to this question, they are guaranteed to break part of Poincaré symmetry in general. Indeed some of the generators explicitly broken by this new family of states (those not in the subalgebra $so(1,d) \in so(2,d)$ associated to de Sitter isometries) are mapped to generators of the Poincaré algebra $iso(1,d-1)$. I thank Ben Craps for discussing this point.} We expect that a complete description of their ultraviolet properties, in line with the issues raised above, can be achieved with better mathematical control than in generic (non-holographic) QFTs. We leave this open for future works.\\ 

\begin{figure}[h!] 
	\centering
	\begin{tikzpicture}
	\draw (-2,3.5) node {\textsc{CFT in} dS\textsubscript{d}};
	\draw (6.2,3.5) node {\textsc{classical fields in} AdS\textsubscript{d+1}};
	\draw[<->] (0,2.5) -- (3,2.5);
	\draw (-2.8,2.5) node {Conformal group SO$(2,d)$};
	\draw (5.9,2.5) node {Isometry group SO$(2,d)$};
	\draw[<->] (0,1.5) -- (3,1.5);
	\draw[->] (-2,1) -- (-2,0.5);
	\draw[->] (5.5,1) -- (5.5,0.5);
	\draw (-2,1.5) node {Sources $\mathcal{J}_{\Delta_i}$};
	\draw (5.5,1.5) node {Boundary conditions};
	\draw[<->] (0,0) -- (3,0);
	\draw (5.5,0) node {Massive fields $\{\phi_{m_i}\}$};
	\draw (-2.7,0) node {Primary operators $\{ \mathcal{O}_{\Delta_i}\}$};
	\draw (1.5,0.5) node {\scalebox{0.7}{$\Delta_i=\frac{d}{2}\pm\sqrt{\frac{d^2}{4}+m_i^2}$}};
	\draw[<->] (0,-1.5) -- (3,-1.5);
	\draw (6.3,-1.5) node {Choice of positive frequencies};
	\draw (-2.3,-1.5) node {Quantum state $|\alpha_\mu \rangle$};
	\draw[->] (5.5,-1) -- (5.5,-0.5);
	\draw[->] (-2,-1) -- (-2,-0.5);
	\end{tikzpicture}
	\caption{Holographic dictionary.}
	\label{Fig:dictionary}
\end{figure}
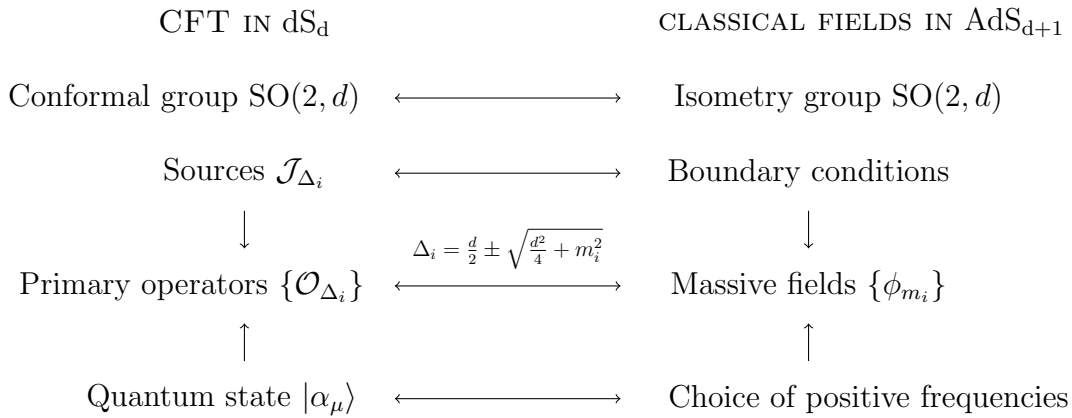

In the considered holographic CFTs, a primary operator $\mathcal{O}_\Delta$ is dual to a classical massive field $\phi_{m_\Delta}$ in anti-de Sitter (AdS). As is it well-known, de Sitter-invariant states in free massive theories may be specified by an appropriate choice of initial positive frequency modes which is equivalent to a choice of initial conditions. For holographic CFTs, we extend this approach by looking at positive frequency modes of the dual classical AdS fields and discover a whole family of de Sitter-invariant CFT states. We call them $\alpha_\mu$-\textit{states} as they are characterized by a continuous set of complex parameters $\alpha_\mu \in \mathbb{C}$ (one for each value of $\mu \in \mathbb{R}_+$). The holographic dictionary, which is reviewed along the text, is schematically represented in Figure \ref{Fig:dictionary}. Strictly speaking, CFTs with classical holographic duals have an infinite number $N$ of degrees of freedom. Computations of finite $1/N$ corrections require to study quantum effects in the AdS dual theory. Here we restrict our study to the limit $N \to \infty$ for which the AdS dual theory is purely classical and assume that some of the CFT degrees of freedom are described by scalar primary operators $\mathcal{O}_\Delta$. Along the text we also make some comments on our expectations for the first subleading $1/N$ corrections.\\

In Section \ref{section:alphavacua} we analyze the wave equation of a free massive scalar field $\phi$ in AdS\textsubscript{d+1} dual to some primary operator in the boundary unitary CFT, and show that it can be decomposed on basis modes of the form\footnote{A factorization of the AdS wave equation has also been uncovered in the context of the `bubble of nothing', leading to similar discussions of inequivalent vacua \cite{Balasubramanian:2003is,Hutasoit:2009sc}.}
\begin{equation}
\label{intro}
\phi_{\vec{k},\lambda}(r,n)=\psi_\lambda(r)\chi_{\vec{k},\lambda}(n), \qquad \vec{k}\in \mathbb{R}^{d-1}, \quad \lambda \in \left[\frac{(d-1)^2}{4},\infty \right[,
\end{equation}
where $\psi_\lambda$ solves a radial Sturm-Liouville problem and $\chi_{\vec{k},\lambda}$ is a de Sitter plane wave with effective mass $m_\lambda^2=\lambda$. Very interestingly, all masses from the \textit{principal series}, $m_\lambda^2 \geq (d-1)^2/4$, appear in this basis. On the other hand modes from the \textit{complementary series}, $m_\lambda^2< (d-1)^2/4$, are forbidden as their associated radial part $\psi_\lambda$ is not normalizable in AdS. Holographic CFTs clearly select well-behaved oscillatory modes of the principal series rather than exponentially growing/decaying ones from the complementary series. Similar in spirit to the Mottola-Allen construction, appropriate choices of positive frequency modes for the dual AdS fields will lead to CFT ``$\alpha_\mu$-states'' preserving de Sitter isometries. This invariance will be checked at the level of two-point correlators of primary operators. Earlier computations of two-point correlators in the unique conformal vacuum may be found in \cite{Koyama:2001rf,Hinterbichler:2015pta,Chu:2016uwi}.\\ 

As we describe in Section \ref{section:spectralbreaking}, expressions for two-point correlators in $\alpha_\mu$-states obtained in Section \ref{section:alphavacua} take the form of spectral decompositions on a basis of free massive de Sitter propagators of the principal series.\footnote{A similar structure had been uncovered in the flat boundary case \cite{Satoh:2002bc}.} We therefore review in Section \ref{section:spectralbreaking} the spectral decomposition of Euclidean two-point correlators studied by Bros \textit{et al}\cite{Bros:1995js} and relate it to a hypergeometric transform known in the mathematics literature as \textit{Olevsky index transform} \cite{Yaku,Neretin}. We explain how our results generalize the spectral decomposition to two-point correlators evaluated in $\alpha_\mu$-states. Those encode non-trivial initial conditions and therefore only exist in a Lorentzian theory. Using properties of the spectral decomposition, \textit{explicit breaking of conformal symmetry} by $\alpha_\mu$-states will be apparent. As an aside, we emphasize that the Olevsky transform may provide computational and mathematical power in wider studies of de Sitter QFTs.\\ 
   
In Section \ref{section:universal} we comment on two other CFT observables characterizing $\alpha_\mu$-states : the stress-energy tensor expectation value and the entanglement entropy associated to cosmological horizons. In the limit $N \to \infty$, their dual representation is purely geometrical such that they take universal values among the family of $\alpha_\mu$-states. We argue that this is in line with existing independent results applying to the conformally coupled scalar \cite{Kanno:2014lma}. However, one expects $1/N$ corrections to distinguish between $\alpha_\mu$-states. Earlier holographic computations of de Sitter entanglement entropies may be found in \cite{Hawking:2000da,Fischler:2013fba,Maldacena:2012xp}.\\

In Section \ref{section:dictionary} we clarify some formal aspects of the dictionary. In particular, we give the correspondence between bulk and boundary operators, both in position and momentum space. Finally,  we comment on a formal interpretation of $\alpha_\mu$-states as the resulting \textit{squeezed states} obtained by turning on sources for the \textit{composite operator} $\mathcal{O}_{\Delta} \mathcal{O}_{\Delta}$, in the conformal vacuum. These formal expressions must however be taken with caution, as a UV-regularization is likely to be required once finite $1/N$ corrections are considered.\\

We hope that the construction presented here can be used as a basis for further studies of holographic CFTs in de Sitter space, including those effects associated to $\alpha_\mu$-states. In this work we simply consider free scalar fields in the bulk of AdS, which are basic ingredients needed to study any specific holographic CFT. Although we focus on states preserving de Sitter isometries, most of the technology developped here can be used to describe completely generic initial holographic CFT states, specified by choices of positive frequency modes in AdS. Computation of subleading $1/N$ corrections would be an interesting avenue for future investigation. These would require a treatment of perturbative quantum effects in the AdS dual theory, such that we expect the framework of time-dependent interacting QFT in curved spacetime to be of important relevance \cite{Schalm:2004xg,Collins:2005cm}. See Sections \ref{section:universal} and \ref{section:dictionary} for more detailed expectations of the first subleading $1/N$ corrections.\\   

Most of the technical details and computations are relagated to appendices where further useful information may be found.

\section{Construction of invariant CFT states} \label{section:alphavacua}
Let's first describe the geometric set-up used throughout this work. The $d$-dimensional de Sitter spacetime dS\textsubscript{d} is most easily described by embedding in $(d+1)$-dimensional Minkowski space $\mathbb{M}^{1,d}$ with the constraint equation
\begin{equation}
\label{dS}
n^2=-(n^0)^2+(n^k)^2=1, \quad k=1,...,d.
\end{equation}
From this it is obvious that its isometry group is the pseudorotation group SO(1,$d$). Denoting by $l$ the de Sitter curvature radius, its line element is induced from $\mathbb{M}^{1,d}$:
\begin{equation}
ds^2_{dS}\equiv l^2dn^2.
\end{equation}
Similarly, AdS\textsubscript{d+1} with unit curvature radius can be embedded in $\mathbb{M}^{2,d}$ through the constraint equation  
\begin{equation}
\eta_{MN}X^{M}X^{N}=-X_0^2+X_1^2+...+X_d^2-X_{d+1}^2=-1, \qquad M,N=0,...,d+1.
\end{equation}
This embedding in $\mathbb{M}^{2,d}$ is very natural in the holographic context as there is a direct connection to the CFT embedding formalism \cite{Costa:2011mg}. We identify a conformal boundary of AdS\textsubscript{d+1} as a section of the light-cone in $\mathbb{M}^{2,d}$ defined through
\begin{equation}
\eta_{MN}P^M P^N=0.
\end{equation}
Since we are interested in CFT in de Sitter space, we introduce a foliation of AdS\textsubscript{d+1}, valid for $X_{d+1}>1$, in terms of $d$-dimensional de Sitter `slices' and we choose a specific section $\mathcal{B}_d$ of the light-cone as anti-de Sitter conformal boundary:
\begin{align}
\label{dS_slicing}
X^M&=\left(\sinh r\ n^\alpha,\cosh r\right) \in AdS_{d+1}, \qquad \alpha=0,...,d,\\
P^M&=l\left(n^\alpha,1\right) \in \mathcal{B}_d,
\end{align}
where $n^\alpha$ satisfies the de Sitter constraint equation \eqref{dS}. Similarly to AdS/CFT with flat boundary \cite{Costa:2011mg}, the limit $r\to \infty$ is interpreted as the location of the conformal boundary $\mathcal{B}_d$ since $X^M \to \cosh r\ P^M$ in this limit. The associated metrics on this patch and on the conformal boundary are
\begin{align}
\label{slicedmetric}
ds^2_{AdS}&= dr^2+\sinh^2 r\ dn^2,\\
\label{boundarymetric}
dP^2&=l^2 dn^2=ds^2_{dS}.
\end{align}
In order to use the holographic dictionary, we fix the value of the de Sitter curvature radius to $l=1/2$. This comes naturally when expressing the AdS metric \eqref{slicedmetric} in Fefferman-Graham gauge, $\rho\equiv e^{-2r}$,
\begin{align}
\label{Fefferman}
ds^2_{AdS}&=\frac{d\rho^2}{4\rho^2}+\frac{(1-\rho)^2}{4\rho} dn^2 \equiv \frac{d\rho^2}{4\rho^2}+\frac{g_{ij}(\rho,n)}{\rho}\ dn^idn^j,
\end{align}
and identifying the boundary metric on $\mathcal{B}_d$ as $\lim\limits_{\rho\to 0} g_{ij}(\rho,n)\equiv g_{(0)ij}(n)$.
    
\subsection{Dual description in AdS} \label{subsection:WaveEquation}
As briefly exposed in the introduction, primary operators in holographic CFTs are dual to classical fields in anti-de Sitter space. For simplicity, we will restrict here to free massive scalar fields. We start by analyzing the Klein-Gordon equation for one such classical field,
\begin{equation}
\label{KG1}
\left(\square_{AdS}-m^2\right) \phi=0,
\end{equation}
and recall that the problem naturally factorizes into a radial Sturm-Liouville problem and a Klein-Gordon equation on the dS\textsubscript{d} slices \cite{Andrade:2011nh}.\footnote{See also \cite{Bertola:2000mx} for similar decompositions in closely related geometries.} A similar structure was also uncovered in the AdS-CFT correspondence with flat boundary \cite{Satoh:2002bc}. Using the foliation \eqref{dS_slicing}, equation \eqref{KG1} becomes
\begin{align}
\left[\sinh^2 r\ \partial_r^2 + d \sinh r\ \cosh r\ \partial_r -\sinh^2 r\ m^2+\square_{dS}\right]\phi=0.
\end{align}
Applying variable separation 
\begin{equation}
\phi(r,n)=\psi(r)\chi(n),
\end{equation}
we then have to solve the eigenvalue problems 
\begin{align}
\label{SL}
\left[\sinh^2 r\ \partial_r^2 + d \sinh r\ \cosh r\ \partial_r -\sinh^2 r\ m^2\right]\psi(r)=-\lambda \psi(r),
\end{align}
\begin{equation}
\label{KG2}
\left(\square_{dS}-\lambda\right) \chi(n)=0.
\end{equation}
Given solutions $\psi_\lambda, \chi_\lambda$ with eigenvalue $\lambda$, the general solution to \eqref{KG1} is constructed as linear combination $\sum_{\lambda}c_\lambda \psi_\lambda(r) \chi_\lambda(n)$. One sees that \eqref{KG2} is nothing but the de Sitter Klein-Gordon equation for a free scalar field of mass $m_\lambda^2=\lambda$. Let's proceed by first solving the radial equation \eqref{SL}. This is a Sturm-Liouville problem on $\mathbb{R}_+$ with weight $w(r)=\sinh^{d-2} r$. Performing the variable change $z=\cosh r \in \left]1,\infty\right[$ and defining
\begin{align}
\psi(z)&\equiv(z^2-1)^{\frac{1-d}{4}}\bar{\psi}(z),\\
\label{mu}
\mu&\equiv\sqrt{\lambda-\frac{(d-1)^2}{4}},\\
\nu&\equiv \sqrt{\frac{d^2}{4}+m^2},
\end{align}
equation \eqref{SL} becomes a Legendre equation:
\begin{equation}
\left[(1-z^2)\partial_z^2-2z\partial_z+\left(m^2+\frac{d^2-1}{4}\right)+\frac{\mu^2}{1-z^2}\right]\bar{\psi}(z)=0.
\end{equation}
Normalizable solutions (in the distributional sense) exist for $\mu \in \mathbb{R}_+$. According to the value of the scalar field mass and following the standard discussion of \cite{Klebanov:1999tb}, we treat two main cases. First note that stability of the scalar field theory requires $m^2 \geq m^2_{BF}\equiv -\frac{d^2}{4}$ \cite{Breitenlohner:1982bm}, which we assume from now on. Solutions corresponding to Dirichlet conditions on $\phi$ at the conformal boundary $\mathcal{B}_d$ are normalizable,
\begin{align}
\label{radialfunction1}
\psi_\mu(z)=\frac{\sqrt{\sinh\mu\pi}e^{-\mu\pi}}{\pi}\ (z^2-1)^{\frac{1-d}{4}}\ Q_{\nu-\frac{1}{2}}^{-i\mu}(z).
\end{align}
Here $Q^a_b(z)$ is the associated Legendre function of the second kind. The scaling dimension of the dual CFT primary operator is 
\begin{equation}
\label{scaling}
\Delta_+\equiv\frac{d}{2}+\nu\geq \frac{d}{2}.
\end{equation}
In the Breitenlohner-Freedman window $m^2_{BF}\leq m^2\leq m^2_{BF}+1$, solutions corresponding to Neumann boundary conditions are also normalizable,
\begin{align}
\label{radialfunction2}
\psi_\mu(z)=\frac{\sqrt{\sinh\mu\pi}e^{-\mu\pi}}{\pi}\ (z^2-1)^{\frac{1-d}{4}}\ Q_{-\nu-\frac{1}{2}}^{-i\mu}(z).
\end{align}
In this case, the scaling dimension of the associated primary operator is 
\begin{equation}
\label{scaling2}
\Delta_-\equiv\frac{d}{2}-\nu\geq \frac{d}{2}-1.
\end{equation}
These normalizable solutions allow one to describe scalar primary operators $\mathcal{O}_\Delta$ of arbitrary scaling dimension above the unitarity bound $\Delta_{unitary}\equiv \frac{d}{2}-1$, as can be seen from \eqref{scaling} and \eqref{scaling2}. A study of CFT operators with scaling dimensions below this bound can be found in \cite{Andrade:2011nh}. To derive the proper normalization constant in \eqref{radialfunction1} and \eqref{radialfunction2}, it has been necessary to make use of the orthogonality relation between solutions of this Sturm-Liouville problem. Although it is guaranteed to exist following from general Sturm-Liouville theory, the orthogonality relation associated to this specific problem was not known so far. We derive it in Appendix \ref{Appendix:Legendre}. Let's turn to the second eigenvalue problem \eqref{KG2}, which is the de Sitter wave equation for a massive scalar field whose mass belongs to the \textit{principal series}, $m^2_\lambda\geq (d-1)^2/4$. One can solve \eqref{KG2} in many coordinate systems and essentially all known results from free massive scalar fields in de Sitter spacetime may be used. Of particular interest is the existence of inequivalent vacua associated to distinct choices of initial positive frequency modes. See Appendix \ref{Appendix:freeQFT} for a review. One basis of positive frequency modes solutions to \eqref{KG1} is given by
\begin{equation}
\label{modesolution}
\phi^{E}_{\mu,\vec{k}}(z,n)=\psi_\mu(z)\chi^{E}_{\mu,\vec{k}}(n),
\end{equation}
which we label by the de Sitter wave vector $\vec{k}$ in analogy to \eqref{modes}, and by the mass parameter $\mu$. The expression of the Euclidean/Bunch-Davies mode functions $\chi_{\mu,\vec{k}}^E$ is given in \eqref{euclidean}. Moreover one can check that these modes are correctly normalized with respect to the Klein-Gordon inner product:
\begin{align}
\label{orthonormality}
\left(\phi^E_{\mu_1,\vec{k}_1},\phi^E_{\mu_2,\vec{k}_2}\right)\equiv -i \int_{\Sigma} d\Sigma^\mu\ \phi^E_{\mu_1,\vec{k}_1}(X) \overleftrightarrow{\partial_\mu} \phi^E_{\mu_2,\vec{k}_2}(X)^*=\delta(\lambda_1-\lambda_2)\ \delta^{d-1}(\vec{k}_1-\vec{k}_2),
\end{align}
where $\Sigma$ is any spacelike hypersurface. A construction similar to the Mottola-Allen one is now possible. We define other bases of positive frequency modes,
\begin{equation}
\label{factorization}
\phi^{(\alpha_\mu)}_{\mu,\vec{k}}(z,n)=\psi_\mu(z)\chi^{(\alpha_\mu)}_{\mu,\vec{k}}(n),
\end{equation}
from Bogolyubov transformations of the form\footnote{\label{note} Note that in earlier literature a different normalization has been used. However the normalization presented here leads to the correct Feynman and retarded propagators needed to satisfy the Klein-Gordon equation \eqref{KG} with normalized $\delta^d(n-n')$ source. This view has also been adopted by the authors of \cite{Anous:2014lia}.} \cite{Spradlin:2001nb}
\begin{equation}
\label{Bogolyubov}
\chi_{\mu,\vec{k}}^{(\alpha_\mu)}\equiv \frac{1}{\sqrt{1+e^{\alpha_\mu+\alpha_\mu^*}}} \left(\chi_{\mu,\vec{k}}^E+e^{\alpha_\mu} \chi_{\mu,-\vec{k}}^{E*}\right), \qquad \text{Re}(\alpha_\mu)\leq 0.
\end{equation}
One may guess that the choice $\chi^{(\alpha_\mu)}_{\mu,\vec{k}}=\chi^E_{\mu,\vec{k}}$ will lead to the unique conformal vacuum, and that other choices defined from \eqref{Bogolyubov} will lead to de Sitter-invariant states. We will check in Section \ref{subsection:Twopoint} that this intuition is correct by looking at two-point correlators which will be manifestly invariant under de Sitter isometries. For obvious reasons, we will refer to the associated de Sitter-invariant CFT states as $\alpha_\mu$-\textit{states}. We emphasize that the main difference with the Mottola-Allen construction \cite{Chernikov:1968zm,Mottola:1984ar,Allen:1985ux} is that we are looking at a scalar primary operator $\mathcal{O}_\Delta$ of arbitrary scaling dimension $\Delta$ and that the choice of positive frequency modes (initial conditions) is made at the level of the dual massive field $\phi$ in AdS. Thanks to the factorization \eqref{factorization} where free massive de Sitter modes $\chi_{\mu,\vec{k}}$ explicitly appear in the dual AdS theory, we still retain much of our intuition about de Sitter physics. At a technical level, one has to treat modes from the entire principal series instead of the subgroup associated to some fixed mass. This means that for the Bogolyubov transformation \eqref{Bogolyubov} to be defined, a complex parameter $\alpha_\mu$ must be specified for each value $\mu \in \mathbb{R}_+$.\\

We end this subsection with  a derivation of \textit{bulk-to-bulk propagators} of the AdS scalar field $\phi$. Obtaining these is essentially equivalent to solving the theory. While in the Euclidean continuation of anti-de Sitter (EAdS) there is a unique propagator, this is completely different in Lorentzian signature. From a mathematical point of view, the reason is that there are normalizable waves in AdS that are square integrable near the boundary. Therefore there exists an infinity of AdS propagators that differ by linear superpositions of these normalizable modes. This does not happen in EAdS as there are no such modes. From a more physical point of view this variety of bulk propagators should reflect the variety of propagators in the boundary CFT such as Wightman, Feynman, retarded, as well as the existence of non-equivalent vacua of the theory \cite{Balasubramanian:1998de,Son:2002sd,Balasubramanian:2003is,Skenderis:2008dg}. Bulk-to-bulk Wightman propagators are constructed as\footnote{This expression  may be alternatively obtained by quantizing the free scalar field $\phi$ in AdS and computing $\langle \alpha_\mu| \hat{\phi}(X) \hat{\phi}(X')|\alpha_\mu \rangle$. We recall that a the level of free fields in a fixed AdS background, the quantized theory contains almost the same information as the classical theory.}
\begin{subequations}
\begin{align}
G_{\alpha_\mu}^{AdS}(X,X')&=\int_{\frac{(d-1)}{4}^2}^\infty d\lambda\ \int d^{d-1}\vec{k}\  \phi^{(\alpha_\mu)}_{\mu,\vec{k}}(X) \phi^{(\alpha_\mu)}_{\mu,\vec{k}}(X')^*\\
&=\int_0^\infty d\mu\ 2\mu\ \psi_\mu(r)\psi_\mu(r')^* \int d^{d-1}\vec{k}\ \chi^{(\alpha_\mu)}_{\mu,\vec{k}}(n) \chi^{(\alpha_\mu)}_{\mu,\vec{k}}(n')^*\\
\label{modesum}
&=\int_0^\infty d\mu\ 2\mu\ \psi_\mu(r)\psi_\mu(r')^*\ G_{\alpha_\mu}^{dS}(n,n';\mu).
\end{align}
\end{subequations}
where $G_{\alpha_\mu}^{dS}(n,n';\mu)$ has been recognized as the Wightman propagator of a free massive scalar field in de Sitter space with mass $m_\lambda^2=\lambda$ (see Appendix \ref{Appendix:freeQFT}). For completeness we also present \textit{boundary-to-bulk propagators} in Appendix \ref{Appendix:Boundary-bulk}. They are useful holographic ingredients for computation of higher $n$-point correlators \cite{Freedman:1998tz}.

\subsection{Two-point correlators} \label{subsection:Twopoint}
What is the quantity dual to the Wightman function $G_{\alpha_\mu}^{AdS}(X,X')$ in the boundary CFT? A first point to note is that it should also be labeled by a complex function $\alpha_\mu$. Hence let us naively denote $|\alpha_\mu \rangle$ the associated CFT state in the boundary theory, such that each choice of function $\alpha_\mu$ would correspond to a different $\alpha_\mu$-\textit{state}.\footnote{Similar reasonings have appeared in various other holographic correspondences \cite{Bousso:2001mw,Spradlin:2001nb,Balasubramanian:2003is,Hutasoit:2009sc}.} In Section \ref{section:dictionary} we formally identify them as \textit{squeezed states} over the unique conformal vacuum. Then, observables in the CFT such as Wightman two-point functions are computed from the renormalized on-shell action of the AdS field $\phi$ \cite{Skenderis:2008dg} or alternatively through the extrapolate dictionary \cite{Banks:1998dd,Harlow:2011ke}. We simply give the result:
\begin{align}
\label{two-point}
\langle \alpha_\mu|\mathcal{O}_\Delta(n_1) \mathcal{O}_\Delta(n_2)|\alpha_\mu\rangle=-4i \nu^2 \lim\limits_{r_1,r_2 \to \infty} (e^{r_1}e^{r_2})^\Delta\ G^{AdS}_{\alpha_\mu}(X_1,X_2).
\end{align}
If one wants to compute other correlators such as Feynman or retarded ones, the procedure is completely transparent: one should have corresponding states and time-orderings on both sides of \eqref{two-point}. For definiteness we focus our analysis on Wightman functions. Using \eqref{radialfunction1} (or \eqref{radialfunction2} if $\Delta$ is in the Breitenlohner-Freedman window) and \eqref{modesum}, equation \eqref{two-point} is reduced to
\begin{equation}
\label{two-point2}
\scalebox{0.99}{$\langle \alpha_\mu| \mathcal{O}_\Delta(n_1) \mathcal{O}_\Delta(n_2)| \alpha_\mu \rangle=-\frac{i2^{d+2}}{\pi\Gamma\left[\nu\right]^2} \int_0^\infty d\mu\ \mu \sinh\mu\pi\ \Big|\Gamma\left[\nu+i\mu+\frac{1}{2}\right]\Big|^2 G_{\alpha_\mu}^{dS}(n_1,n_2;\mu).$}
\end{equation}  
These are manifestly \textit{invariant under de Sitter isometries}, since the massive propagators $G_{\alpha_\mu}^{dS}(n_1,n_2;\mu)$ satisfy themselves this property (see Appendix \ref{Appendix:freeQFT}), and we conclude that $\alpha_\mu$-states are de Sitter-invariant. Perhaps more interestingly, equation \eqref{two-point2} expresses the two-point function of a primary operator as linear superposition of free de Sitter propagators of masses from the \textit{principal series}. This is usually referred to as a spectral decomposition. In the context of de Sitter QFTs, spectral decomposition properties of Euclidean propagators have been studied in \cite{Bros:1995js}. In the next section we point out that it is simply related to the \textit{Olevsky index transform}. Using this transform, we will argue that $\alpha_\mu$-states explicitly break conformal invariance (except for the conformal vacuum).

\section{Spectral decomposition and breaking of conformal symmetry} \label{section:spectralbreaking}
\subsection{Spectral decomposition and  Olevsky transform} \label{subsection:spectral}  
It was shown by Bros \textit{et al} \cite{Bros:1995js} that Euclidean Wightman functions $W_E(n,n')$ of any de Sitter QFT can be decomposed over the \textit{principal series}, $\mu \in \mathbb{R}_+$, of free massive Euclidean Wightman functions $G_E^{dS}(n,n';\mu)$:\footnote{A simple way to understand this decomposition theorem is by noting that free Wightman functions $G^{dS}(n,n';\mu)$ are solutions to the Sturm-Liouville eigenvalue problem  
	\begin{equation}
	\nonumber
	\left[\frac{d}{dx}\left(x(1+x)\right)^{\frac{d}{2}}\frac{d}{dx}\right] G(x)=-\lambda w(x) G(x), \qquad w(x)\equiv\left(x(1+x)\right)^{\frac{d}{2}-1},
	\end{equation}
	where $x\equiv -\frac{1+n.n'}{2}$ and the eigenvalue is related to the field mass by $m^2_\lambda=\lambda$. This is actually nothing else than the Klein-Gordon equation written in terms of the two-point invariant variable $x$. If we impose as boundary conditions on generic Wightman functions to belong to $L^2\left[\mathbb{R}_+, w(x)dx\right]$, then the Euclidean propagators $G^{dS}_E(n,n';\mu)$  of the principal series $\mu \in \mathbb{R}_+$ form a basis for such functions. Other choice of boundary conditions would select a linear combination of $G^{dS}_E(x;\mu)$ and $G^{dS}_E(1-x;\mu)$ as natural basis, such as $G^{dS}_{\alpha_\mu}(x;\mu)$. Be careful though that one should always ensure square integrability $L^2\left[\mathcal{D}, w(x)dx\right]$ on some domain $\mathcal{D}$ of interest.
	\label{footspectral}}
\begin{align}
\label{spectral}
W_E(n,n')=\int_0^\infty d\mu\ \rho(\mu)\ G_E^{dS}(n,n';\mu).
\end{align}
Moreover it was shown that the weight function $\rho(\mu)$ must be  positive-definite. This equation is the de Sitter analogue of the Källen-Lehmann spectral decomposition in flat space.\\

Let's connect this decomposition with the so-called Olevsky transform \cite{Yaku,Neretin}. This is a unitary operation from $L^2\left[\mathbb{R}_+, x^{a-1}(1+x)^{a-2b}dx\right]$ to $L^2\left[\mathbb{R}_+, |\Gamma\left[a-b+i\mu\right]/\Gamma\left[b+i\mu\right]|^2 d\mu\right]$ for any $a>b>0$. The transform and its inverse are given by
\begin{align}
\label{Olevsky}
\left(\mathcal{T}f\right)(\mu)\equiv\hat{f}(\mu)\equiv \frac{|\Gamma\left[b+i\mu\right]|^2}{\Gamma\left[a\right]} \int_0^\infty dx\ x^{a-1} {}_2F_1\left[b+i\mu,b-i\mu;a;-x\right] f(x),
\end{align}
\begin{equation}
\label{inverseOlevsky}
\scalebox{0.97}{$f(x)=\frac{1}{\pi^2 \Gamma\left[a\right]}\int_0^\infty d\mu\ \mu \sinh 2\pi\mu\ |\Gamma\left[a-b+i\mu\right]|^2\ {}_2F_1\left[a-b+i\mu,a-b-i\mu;a;-x\right] \hat{f}(\mu).$}
\end{equation}
\newline
Using expression \eqref{Twopointgauss} for the free massive Wightman function $G_E^{dS}(n,n';\mu)$, it is straightforward to see that the spectral decomposition \eqref{spectral} is actually the inverse Olevsky transform of $G^{dS}_E(n,n';\mu)$ as defined in \eqref{inverseOlevsky} with $x=-\frac{1+n.n'}{2}$, $a=\frac{d}{2}$, $b=\frac{1}{2}$ and $\rho(\mu) \sim \mu \sinh 2\pi\mu\ \hat{f}(\mu)$.\\

Coming back to the holographic correspondence, we see that the holographic formula \eqref{two-point2} is the natural extension of the spectral decomposition \eqref{spectral} to $\alpha_\mu$-states Wightman functions : those decompose over corresponding Mottola-Allen Wightman functions $G_{\alpha_\mu}^{dS}(n,n';\mu)$. 

\subsection{The conformal vacuum} \label{subsection:conformalvacuum}
A two-point function satisfying conformal invariance has the form
\begin{align}
\label{Twopointconformal2}
\langle \mathcal{O}_{\Delta}(n) \mathcal{O}_{\Delta}(n') \rangle=C_{\Delta} \left(\frac{2}{1-n.n'}\right)^{\Delta},
\end{align}
Details of the derivation may be found in Appendix \ref{Appendix:CFT}. We omit $i\epsilon$ insertions but these can be reinstated easily from the discussion found in Appendix \ref{Appendix:freeQFT}. This two-point function has a well-defined Olevsky transform, meaning that we can decompose it on Euclidean propagators $G_E^{dS}(n,n';\mu)$. Let's find the associated spectral weight, which amounts to performing its Olevsky transform. Using Lemma 2.1 of \cite{Neretin} and formula 6.412 of \cite{Gradshteyn}, we get
\begin{subequations}
\label{twopointtransform}
\begin{align}
\nonumber
&\left(\mathcal{T}\langle \mathcal{O}_\Delta(n)\mathcal{O}_\Delta(n') \rangle \right)(\mu)\\
&=C_{\Delta}\ \frac{|\Gamma\left[\frac{1}{2}+i\mu\right]|^2}{\Gamma\left[\frac{d}{2}\right]} \int_0^\infty dx\ \frac{x^{d/2-1}}{(1+x)^{\Delta}}\ {}_2F_1\left[\frac{1}{2}+i\mu,\frac{1}{2}-i\mu;\frac{d}{2};-x\right],\\
&=\frac{1}{2\pi i} \frac{C_{\Delta}}{\Gamma\left[\Delta\right]} \int_{-i\infty}^{i\infty} ds\ \Gamma\left[\Delta-\frac{d}{2}-s\right]\Gamma\left[\frac{1}{2}+i\mu+s\right]\Gamma\left[\frac{1}{2}-i\mu+s\right]\Gamma\left[-s\right]\\
&=C_{\Delta}\ \frac{\Big|\Gamma\left[\frac{1}{2}+i\mu\right]\Big|^2\Big|\Gamma\left[\Delta-\frac{d+1}{2}+i\mu\right]\Big|^2}{\Gamma\left[\Delta\right]\Gamma\left[\Delta-\frac{d-2}{2}\right]}=C_{\Delta}\ \frac{\pi}{\cosh \pi\mu}\frac{\Big|\Gamma\left[-\frac{1}{2}+\nu+i\mu\right]\Big|^2}{\Gamma\left[\Delta\right]\Gamma\left[1+\nu\right]}.
\end{align} 
\end{subequations}
One might have guessed that the choice $\chi_{\mu,\vec{k}}^{(\alpha_\mu)}=\chi_{\mu,\vec{k}}^E\ $ for positive frequency modes would correspond to the \textit{conformal vacuum}. Indeed, with this choice and the use of \eqref{twopointtransform} it is easy to check that equation \eqref{two-point2} reduces to
\begin{align}
\langle E| \mathcal{O}_\Delta(n_1) \mathcal{O}_\Delta(n_2)| E \rangle=-2i\nu \frac{\Gamma\left[\Delta\right]}{\pi^{d/2}\Gamma\left[\nu\right]} \left(\frac{2}{1-n_1.n_2}\right)^{\Delta}, \qquad |E \rangle\equiv|\alpha_\mu=-\infty \rangle.
\end{align}
We conclude that $|E \rangle$ is the unique conformal vacuum. Moreover, since the theory in de Sitter space is related by a conformal transformation to the theory in flat Minkowski space (see Appendix \ref{Appendix:freeQFT}), 
\begin{align}
ds^2_{dS}=l^2 \frac{-d\eta^2+d\vec{x}^2}{\eta^2}=\Omega^2(x)ds^2_{flat}, \qquad \Omega(x)=\frac{l}{\eta}=\frac{1}{2\eta}, 
\end{align} 
correlators in the conformal vacuum should be related through
\begin{equation}
\langle E| \mathcal{O}_\Delta(n_1) \mathcal{O}_\Delta(n_2)| E \rangle=\Omega(x_1)^{-\Delta}\Omega(x_2)^{-\Delta}\langle \mathcal{O}_\Delta(x_1) \mathcal{O}_\Delta(x_2)\rangle_{flat}.
\end{equation} 
One can check that this is indeed the case from the expression of $\langle \mathcal{O}_\Delta(x_1) \mathcal{O}_\Delta(x_2)\rangle_{flat}$ given in \cite{Freedman:1998tz}, including numerical constants.

\subsection{Explicit breaking of conformal symmetry}
We turn to the case where the complex parameters are $\mu$-independent and distinct from the conformal vacuum value, $\alpha_\mu=\gamma\neq -\infty$ for all $\mu \in \mathbb{R}_+$. Using \eqref{alphaWightman}, expression of the two-point function \eqref{two-point2} reduces to
\begin{IEEEeqnarray}{l}
\langle \gamma| \mathcal{O}_\Delta(n_1) \mathcal{O}_\Delta(n_2)| \gamma \rangle=-2i\nu \frac{\Gamma\left[\Delta\right]}{\pi^{d/2}\Gamma\left[\nu\right]}\left[\left(\frac{2}{1-n_1.n_2}\right)^{\Delta}+\frac{e^\gamma+e^{\gamma^*}}{1+e^{\gamma+\gamma^*}}\left(\frac{2}{1+n_1.n_2}\right)^{\Delta}\right].
\IEEEeqnarraynumspace
\end{IEEEeqnarray}
The last term does not satisfy the matter conformal Ward identities presented in Appendix \ref{subsection:Ward} and we conclude that generic $\alpha_\mu$-states explicitly break conformal invariance. This was expected since we knew from Section \ref{subsection:conformalvacuum} that conformally invariant two-point functions should be inverse Olevsky transform of \eqref{twopointtransform}, meaning that their spectral decomposition should have $G_E^{dS}(n_1,n_2;\mu)$ as kernel. It was already clear from \eqref{two-point2} that this is not the case when $\alpha_\mu \neq -\infty$. We point out that breaking of conformal symmetry by $\alpha$-vacua has also been observed in the context of $\mathcal{N}=1$ superconformal Yang-Mills theory on 4-dimensional de Sitter space in \cite{Anous:2014lia}.\\

We also consider a case where $\alpha_\mu$ is a function of $\mu$. One interesting such example is an adaptation to the case of holographic CFTs, of \textit{in}/\textit{out} vacua introduced in \cite{Mottola:1984ar} and studied in \cite{Bousso:2001mw} in connection with the dS/CFT duality. We thus define CFT \textit{in}/\textit{out} vacua as $\alpha_\mu$-states characterized by
\begin{equation}
\alpha_\mu=-\pi\mu \pm i\frac{d+1}{2} \pi, \qquad \mu \in \mathbb{R}_+.
\end{equation}
In the context of \cite{Mottola:1984ar} where a free massive scalar field in de Sitter space is considered, $\mu$ takes a single value. As emphasized several times, here we need to assign a complex parameter $\alpha_\mu$ for each value of $\mu \in \mathbb{R}_+$. We have not been able to further reduce \eqref{two-point2} in generality, as it is possibly some unknown distribution. However we can repeat the argument applied to constant $\alpha_\mu$-parameter and conclude that \textit{in}/\textit{out} states must break conformal invariance. For particular values of the scaling dimension $\Delta$ and space dimension $d$ the computation becomes tractable. For example, for $d=3$ and $\Delta=1$ one has\footnote{One should use \eqref{TwopointLegendre} together with equation 8.754.3 of \cite{Gradshteyn}.}
\begin{equation}
\langle in|\mathcal{O}_1(n)\mathcal{O}_1(n')|in\rangle\simeq\left(\frac{1}{1-n.n'}-\frac{1}{1+n.n'} \left(1 - \sqrt{2} (1-n.n')^{-1/2}\right)\right).
\end{equation}
The first contribution is equal to the conformal one, while the second contribution obviously breaks conformal symmetry.\\  

With the use of boundary-to-bulk and bulk-to-bulk propagators, one can go on and compute higher $n$-point correlators from Feynman-Witten diagrams. In Appendix \ref{Appendix:ThreePoint} we illustrate how this works by computing three-point correlators in the conformal vacuum, finding agreement with the flat-space results and conformal symmetry altogether.

\section{Stress-tensor and entanglement entropy} \label{section:universal}
So far we considered correlators of primary fields as observables describing the family of de Sitter-invariant $\alpha_\mu$-states defined implicitly from a choice of positive frequency modes \eqref{Bogolyubov} of the dual AdS fields. We have found that two-point correlators satisfy generalized spectral decompositions and that the holographically dual AdS theory encodes this information in a very natural way. In this section we discuss two other CFT observables : the stress-energy tensor expectation value and the entanglement entropy of the cosmological horizon. In the strict $N \to \infty$ limit, these have in common that their holographic dual is purely geometrical. Indeed, this corresponds to the classical limit of the bulk theory which is described by empty AdS for all $\alpha_\mu$-states. One is led to the conclusion that those observables are universal among the whole family of $\alpha_\mu$-states in that limit. However, one expects that finite $1/N$ corrections should distinguish between $\alpha_\mu$-states. Let's review the status of each two observables separately.\\

The stress-energy tensor expectation value in the classical bulk regime has been computed from holography in full generality in \cite{deHaro:2000vlm} and has been shown to agree with results from conformal field theory. In particular one can see that conformal anomalies are correctly accounted for. For concreteness we present it for de Sitter CFTs of dimension up to $d=4$:
\begin{align}
\label{energytensor1}
d\ \text{odd}: \qquad \langle T_{ij} \rangle&=0,\\
\label{energytensor2}
d=2: \qquad \langle T_{ij} \rangle&=\frac{1}{32\pi G_3}R_{(0)}\ g_{(0)ij},\\
\label{energytensor3}
d=4: \qquad \langle T_{ij} \rangle&=-\frac{1}{3.2^{14}\pi G_5} R_{(0)}^2\ g_{(0)ij}.
\end{align}
Here $g_{(0)ij}$ is the de Sitter boundary metric \eqref{boundarymetric} and $R_{(0)}=d(d-1)/l^2$ is the associated curvature. Since no information about the AdS scalar field $\phi$ is needed in expressions \eqref{energytensor1}-\eqref{energytensor3}, we infer that the stress tensor expectation value is the same in any $\alpha_\mu$-state of the boundary CFT. In free de Sitter QFTs and in particular for the case of a conformally coupled scalar field, Mottola-Allen vacua also share this property as can be seen from equation \eqref{stresstensoralpha} of Appendix \ref{Appendix:freeQFT}. There is therefore nothing surprising with this holographic result in the limit where the bulk theory is classical.\\

Entanglement entropy associated to some bipartition of the boundary theory is computed from the AdS bulk geometry as the minimal surface area whose intersection with the boundary is precisely the bipartition surface \cite{Ryu:2006bv}. In Appendix \ref{Appendix:Entanglement}, we derive the de Sitter entropy arising from tracing out degrees of freedom living behind one inertial  observer's horizon (cosmological horizon). This generalizes the computation of \cite{Hawking:2000da,Li:2011bt} valid for $d=2,4$ to generic dimension $d$. The result exactly reproduces the well-known Gibbons-Hawking formula for de Sitter gravitational entropy \cite{Gibbons:1977mu}:
\begin{equation}
\label{dShorizon}
S=\frac{\text{Horizon}_{\text{dS\textsubscript{d}}}}{4G_d},
\end{equation}
where $\text{Horizon}_{\text{dS\textsubscript{d}}}$ is the area of the cosmological horizon. As for the stress tensor expectation value, this result valid in the limit $N \to \infty$ only depends on the bulk geometry which is empty AdS for all $\alpha_\mu$-states. We point out that entanglement entropies in Mottola-Allen vacua have been computed for free scalar theories in \cite{Kanno:2014lma}. Rather interestingly, the analysis shows that these depend on the vacuum choice except when the mass of the field takes the conformally coupled value. We have demonstrated that this result also holds for de Sitter-invariant states of holographic CFTs in the large $N$ limit.\\

Computation of subleading $1/N$ corrections would require to take into account the change in AdS background geometry due to quantum one-loop contributions to the bulk stress tensor. These one-loop effects originate from quantum fluctuations of all bulk fields including those of the metric field itself. In particular, the state-dependent bulk propagators given in \eqref{modesum} are to be used for computing these one-loop corrections. In turn, the background bulk metric itself would have to adjust such as to satisfy the quantum corrected Einstein equations. Expressions \eqref{energytensor1}-\eqref{energytensor3} for the boundary CFT stress-tensor are sensitive to such a change in the background metric and should in principle distinguish between $\alpha_\mu$-states. The area of minimal surfaces anchored on the boundary is similarly affected and should in particular produce state-dependent corrections to the cosmological horizon entanglement entropy \eqref{dShorizon}. It has been shown in \cite{Faulkner:2013ana} that additional $1/N$ corrections to CFT entanglement entropy arise from one-loop entanglement of bulk fields across the bulk minimal surface, which we also expect to be sensitive to the choice of $\alpha_\mu$-state. As a possible future research avenue, it would be interesting to quantify these $1/N$ corrections.

\section{Aspects of operator and state dictionaries}\label{section:dictionary}
In this section, we clarify some formal aspects of the dictionary. We start by giving the relation between bulk and boundary operators in position and momentum space. The bulk operator can be written as
\begin{align}
\label{bulkoperator}
\hat{\phi}(z,n)=\int_{0}^{\infty} d\mu\ 2\mu \int d^{d-1}\vec{k}\ \left(\phi_{\mu,\vec{k}}^E(z,n)\ \hat{a}_{\mu,\vec{k}}^E + h.c.\right),
\end{align}
where $\hat{a}_{\mu,\vec{k}}^E$ annihilates the bulk representation of the conformal vacuum,
\begin{equation}
\hat{a}_{\mu,\vec{k}}^E |E\rangle=0, \qquad \forall \mu,\vec{k}.
\end{equation}
It is known that one can extract the dual CFT operator $\mathcal{O}_\Delta$ by pushing the bulk operator $\hat{\phi}(z,n)$ to the boundary\cite{Banks:1998dd,Harlow:2011ke},
\begin{equation}
\label{OperatorDictionary}
\mathcal{O}_{\Delta}(n)=2\nu \lim\limits_{z\to 0} z^{\Delta}\ \hat{\phi}(z,n).
\end{equation}
Plugging \eqref{bulkoperator} in \eqref{OperatorDictionary} together with the mode solution \eqref{modesolution}, we get the formula for the annihilation part of $\mathcal{O}_\Delta$, 
\begin{equation}
\mathcal{O}_{\Delta}^{-}(n)=\frac{\nu\ \eta^{\frac{d-1}{2}}}{2^{\nu-\frac{1}{2}}}  \int \frac{d^{d-1}\vec{k}}{(2\pi)^{\frac{d-1}{2}}}\ e^{i\vec{k}.\vec{x}} \int_0^\infty d\mu\ \mu \sqrt{e^{\mu\pi} \sinh \mu\pi}\ H_{i\mu}^{(2)}(k\eta) \frac{\Gamma\left[\nu-i\mu+\frac{1}{2}\right]}{\Gamma\left[\nu+1\right]} \hat{a}_{\mu,\vec{k}}.
\end{equation}
Similarly to what is found in AdS-CFT with flat boundary \cite{Bena:1999jv}, one can obtain the ``momentum space" version of \eqref{OperatorDictionary}, i.e. one can isolate creation and annihilation operators for independent modes by taking suitable transforms:\footnote{The Kontorovich-Lebedev transform and its inverse are given by\cite{Yaku} 
	\begin{align*}
	\left(\mathcal{K}\ f\right)(\mu)&\equiv \int_0^\infty dv\ K_{i\mu}(v)f(v),\\
	f(v)&=\frac{1}{\pi^2 v} \int_0^\infty d\mu\ 2\mu \sinh \pi\mu\ K_{i\mu}(x) \left(\mathcal{K}\ f\right)(\mu),
	\end{align*}
	where $K_{i\mu}$ is the Macdonald function. The derivation of \eqref{operatortransform} requires to use the orthogonality relation
	\begin{equation*} 
	\frac{2}{\pi^2}\int_0^\infty dv\ v^{-1} K_{i\mu}(v) K_{i\mu'}(v)=\frac{\delta(\mu-\mu')}{\mu \sinh \pi\mu}.
	\end{equation*}}
\begin{equation}
\label{operatortransform}
\mathcal{O}^-_\Delta (\mu,\vec{k})\equiv \left(\mathcal{K} \circ \mathcal{F}\ \eta^{-\frac{1+d}{2}} \mathcal{O}^-_{\Delta}\right)(\mu,\vec{k})=-\frac{2\pi\nu}{2^{\nu+\frac{1}{2}}}\frac{|\vec{k}|}{\sqrt{\sinh\ \mu\pi}} \frac{\Gamma\left[\nu-i\mu+\frac{1}{2}\right]}{\Gamma\left[\nu+1\right]}\ \hat{a}_{\mu,\vec{k}}.
\end{equation}
Here we applied a spatial Fourier transform on the operator $\eta^{-\frac{1+d}{2}}\mathcal{O}^-_{\Delta}(\eta,\vec{x})$, followed by a Kontorovich-Lebedev transform with respect to the comoving time variable  $v\equiv e^{i\pi/2} |\vec{k}|\eta$. This relation will help in the following discussion.\\

It is natural to ask how $\alpha_\mu$-states are constructed from the conformal vacuum. Very similarly to the context of Mottola-Allen vacua for free fields in de Sitter \cite{Bousso:2001mw,Einhorn:2003xb}, one can formally write $|\alpha_\mu \rangle$ as a \textit{squeezed state},
\begin{align}
\label{Srotation}
|\alpha_\mu \rangle = \mathcal{U}^{(\alpha_\mu)}|E\rangle,
\end{align}
where
\begin{align}
\label{Smatrix}
\mathcal{U}^{(\alpha_\mu)}&=\exp\left[\int d\mu\ 2\mu \int d^{d-1}\vec{k}\ \left( c(\alpha_\mu)\ \hat{a}_{\mu,\vec{k}}^{E\dagger}\ \hat{a}_{\mu,-\vec{k}}^{E\dagger}-h.c.\right)\right],\\
c(\alpha_\mu)&=\frac{1}{4} \ln\left(\tanh\frac{|\text{Re}\ \alpha_\mu|}{2}\right) e^{-i\text{Im}\ \alpha_\mu}.
\end{align}
Using the operator dictionary \eqref{operatortransform}, one can alternatively find an appropriate representation in the boundary CFT,
\begin{equation}
\label{composite}
|\alpha_\mu \rangle = \exp\left[\int d\mu\ 2\mu \int d^{d-1}\vec{k} \left(\mathcal{J}^{(\alpha_\mu)}(\mu,\vec{k})\ \mathcal{O}^+_\Delta(\mu,\vec{k})\ \mathcal{O}^+_\Delta(\mu,-\vec{k})-h.c.\right)\right]|E\rangle.
\end{equation}   
The expression of $\mathcal{J}^{(\alpha_\mu)}(\mu,\vec{k})$ can be worked out easily. From this formal expression, we deduce that $\alpha_\mu$-states are constructed from the conformal vacuum by sourcing the \textit{composite operators} $\mathcal{O}^+_\Delta(\mu,\vec{k})\ \mathcal{O}^+_\Delta(\mu,-\vec{k})$ and $\mathcal{O}^-_\Delta(\mu,\vec{k})\ \mathcal{O}^-_\Delta(\mu,-\vec{k})$ with sources $\mathcal{J}^{(\alpha_\mu)}(\mu,\vec{k})$ and $-\mathcal{J}^{(\alpha_\mu)}(\mu,\vec{k})^*$, respectively. However one should be cautious in manipulating such formal expressions as one might expect quantum corrections in the AdS bulk theory to require ultraviolet regularization in the spirit of effective field theory, as it has been discussed in the context of Mottola-Allen vacua \cite{Einhorn:2003xb,deBoer:2004nd,Schalm:2004xg}. In this case, formulas \eqref{Smatrix} and \eqref{composite} would only be valid below some ultraviolet cut-off. As already mentioned, these quantum effects in the bulk theory translate into subleading $1/N$ corrections in the boundary CFT, but a precise analysis of these contributions is left for future work.

\section*{Acknowledgments}
I want to thank Ben Craps and Oleg Evnin for careful reading and useful comments on early versions of this manuscript, Vladislav Vaganov for spotting a typo, and the referees of Classical and Quantum Gravity for useful comments on the role of subleading $1/N$ corrections and ultraviolet regularization of Mottola-Allen vacua. I also thank Ben Craps for many structural advices that lead to this final version. This work is supported in part by a PhD fellowship from the VUB Research Council, by the Belgian Federal Science Policy Office through the Interuniversity Attraction Pole P7/37, by FWO-Vlaanderen through project G020714N, and by Vrije Universiteit Brussel through the Strategic Research Program ``High-Energy Physics".

\appendix
\section{Propagators and vacua of free de Sitter QFTs} \label{Appendix:freeQFT}
We review some well-known facts about free massive scalar theories in de Sitter space \cite{Parker,Spradlin:2001pw,Bousso:2001mw}. The defining equation of de Sitter was given in \eqref{dS}. We will mainly make use of the \textit{expanding} (or \textit{conformal}) coordinate patch that covers half of de Sitter:\footnote{The other half can be covered with the same coordinate system and $\eta \in \left]0,\infty\right[$. This is called the \textit{contracting} patch of de Sitter.}
\begin{align}
n^0&=\left(\eta^2-\vec{x}^2-1\right)/2\eta, \qquad \eta \in \left]-\infty,0\right[, \quad \vec{x} \in \mathbb{R}^{d-1},\\
n^d&=\left(\eta^2-\vec{x}^2+1\right)/2\eta,\\
n^i&=\ x^i/\eta, \qquad i=1,...,d-1.
\end{align}
This patch is most suited to discuss conformal theories as it is manifestly conformal to Minkowski,
\begin{equation}
\label{conformalpatch}
ds^2_{dS}\equiv l^2dn^2=\frac{l^2}{\eta^2}\left(-d\eta^2+d\vec{x}^2\right)=\frac{l^2}{\eta^2}\ ds^2_{flat}.
\end{equation}
We denoted the de Sitter curvature radius by $l$. Note that many properties of the conformal vacuum can be obtained by a conformal transformation from Minkowski to the conformal patch of de Sitter. For later use we also give the expression of the invariant two-point variable under SO(1,$d$) isometries:
\begin{equation}
n.n'=\frac{1}{2\eta\eta'}\left(\eta^2+\eta'^2-(\vec{x}-\vec{x}')^2\right).
\end{equation}
Mode solutions of the Klein-Gordon equation with mass $m$ are of the form
\begin{align}
\label{modes}
\chi_{\vec{k}}(n)=\eta^{\frac{d-1}{2}}\left(A H_{i\mu}^{(1)}(k\eta)+BH_{i\mu}^{(2)}(k\eta)\right)\frac{e^{i\vec{k}.\vec{x}}}{\sqrt{2(2\pi)^{d-1}}}, \qquad k\equiv |\vec{k}|,
\end{align}
with $H^{(1)},H^{(2)}$ the Hankel functions of first and second kind and 
\begin{equation}
\mu\equiv \sqrt{m^2-\frac{(d-1)^2}{4}}.
\end{equation}
Notice the analogy with equation \eqref{mu}. Fields with $\mu \in \mathbb{R}_+$ are said to be part of the \textit{principal series}, and associated modes have oscillatory behavior. Fields with imaginary $\mu$ form the \textit{complementary series} and their modes behave exponentially. For a free theory in curved spacetime, definition of a vacuum is equivalent to specification of positive frequency modes associated to annihilation operators. In the present case, one can choose any $\chi_{\vec{k}}(n)$ of the form \eqref{modes} as positive frequency mode, subject to the normalizability condition
\begin{equation}
\left(e^{\mu\pi}AA^*-e^{-\mu\pi}BB^*\right)=-\frac{\pi}{2}.
\end{equation}
Often considered is the \textit{Euclidean} (or \textit{Bunch-Davies}) vacuum defined with respect to the normalized positive frequency modes
\begin{equation}
\label{euclidean}
\chi_{\vec{k}}^{E}(n)=\sqrt{\frac{\pi e^{\mu\pi}}{2}} \eta^{\frac{d-1}{2}} H_{i\mu}^{(2)}(k\eta)\frac{e^{i\vec{k}.\vec{x}}}{\sqrt{2(2\pi)^{d-1}}},
\end{equation}
that behave as $\sim \eta^{d/2-1} e^{-ik\eta}$ in the infinite past $\eta\to- \infty$ of the expanding patch. The Wightman function in this vacuum can be expressed in terms of a Gauss hypergeometric, an associated Legendre or a Gegenbauer function:
\begin{subequations}
\begin{align}
\label{Twopointgauss}
G_E^{dS}(n,n';\mu)\equiv\langle E |\hat{\chi}(n)\hat{\chi}(n')|E\rangle&=\frac{\Gamma\left[h_+\right]\Gamma\left[h_-\right]}{(4\pi)^{\frac{d}{2}}\Gamma\left[\frac{d}{2}\right]} {}_2F_1\left[h_+,h_-;\frac{d}{2};\frac{1+n.n'}{2}\right]\\
\label{TwopointLegendre}
&=\frac{\Gamma\left[h_+\right]\Gamma\left[h_-\right]}{2(2\pi)^{\frac{d}{2}}}((n.n')^2-1)^{\frac{2-d}{4}}P_{-i\mu-\frac{1}{2}}^{1-\frac{d}{2}}(-n.n')\\
\label{TwopointGegenbauer}
&=\frac{\pi\Gamma\left[d-1\right]}{(4\pi)^{\frac{d}{2}}\sin\left(\frac{d-1}{2}+i\mu\right)\Gamma\left[\frac{d}{2}\right]}C^{\frac{d-1}{2}}_{-\frac{d-1}{2}-i\mu}(-n.n'),
\end{align}
\end{subequations}
with $h_{\pm}\equiv\frac{d-1}{2}\pm i\mu$. We have explicitly indexed the Wightman function by the associated field mass parameter $\mu$. The $i\epsilon$ prescription needed for lightlike separated points is $n.n'-i\epsilon \ \text{sgn}(\eta-\eta')$. It is invariant under the connected part of SO(1,$d$) since it depends on the two-point invariant $n.n'$ and the time ordering only. The validity of this expression actually extends to the whole de Sitter space. Moreover one can check that upon analytic continuation to Euclidean signature $\eta = i\eta_E$, it is the only regular solution to the Laplace equation on the sphere $\mathcal{S}^d$.\\

Including the Euclidean vacuum, there is a two-parameter family of vacua invariant under the connected part of SO(1,$d$). These are the Mottola-Allen $\alpha$-vacua and are related to the Euclidean vacuum by a Bogolyubov transformation of complex parameter $\alpha$ of the form \eqref{Bogolyubov}. For a treatment in global coordinates, see \cite{Mottola:1984ar,Allen:1985ux,Bousso:2001mw}. The associated Wightman functions are 
\begin{align}
\label{alphaWightman}
&G_\alpha^{dS}(n,n';\mu)\equiv\langle \alpha | \hat{\chi}(n) \hat{\chi}(n') |\alpha \rangle\\
\nonumber
&=\frac{1}{1+e^{\alpha+\alpha^*}} \left(G_E^{dS}(n,n';\mu)+e^{\alpha+\alpha^*}G_E^{dS}(n',n;\mu)+e^{\alpha^*}G_E^{dS}(n,n_A';\mu)+e^{\alpha}G_E^{dS}(n_A,n';\mu)\right),
\end{align}  
where $n_A=-n$ is the point antipodal to $n$. From this expression, one can see that it is invariant under the connected part of SO(1,$d$). Note that only $\alpha$-vacua with real $\alpha$ are invariant under the full SO(1,$d$) isometry group \cite{Bousso:2001mw}. In contrast to what has been concluded for long in the literature, our choice of normalization (see footnote \ref{note}) leads to a universal singularity at coincident points in the anti-commutator two-point function, irrespective of the $\alpha$ parameter. Indeed in the limit $n\sim n'$,
\begin{subequations}
\begin{align}
&G^{(1)}_\alpha (n,n';\mu)\equiv G_\alpha(n,n';\mu)+G_\alpha(n',n;\mu)\\
&\simeq \frac{1}{e^{\alpha+\alpha^*}} \left(G_E(n,n';\mu)+e^{\alpha+\alpha^*}G_E(n',n;\mu)+G_E(n',n;\mu)+e^{\alpha+\alpha^*}G_E(n,n';\mu)\right)\\
&=G_E(n,n';\mu)+G_E(n',n;\mu)=G^{(1)}_E(n,n';\mu).
\end{align}
\end{subequations}
This has important consequences for the computation of the renormalized stress-energy tensor in these vacua. Using renormalization by point-splitting \cite{Bunch:1978yq}, one deduces that the renormalized stress-energy tensor is the same in any of the $\alpha$-vacua:
\begin{equation}
\label{stresstensoralpha}
\langle \alpha| T_{ij} |\alpha \rangle_{ren}=\langle E| T_{ij} |E\rangle_{ren}\equiv \langle T_{ij} \rangle.
\end{equation}
For completeness we recall that Feynman and retarded Green functions satisfying
\begin{equation}
\label{KG}
\left(\square-m^2\right)G_{F/R}(n,n')=\frac{\delta^d(n-n')}{\sqrt{-g}},
\end{equation}
can be constructed from the Wightman function $G$ as
\begin{align}
iG_F(n,n')&=\theta(t-t')G(n,n')+\theta(t'-t)G(n',n),\\
iG_R(n,n')&=i\theta(t-t')\left(G(n,n')-G(n',n)\right)=2i\theta(t-t')\text{Re}\left[G_F(n,n')\right].
\end{align}

\section{Conformal symmetry in de Sitter space} \label{Appendix:CFT}
In this section we give a basic description of scalar conformal field theories in de Sitter space by applying the standard treatment of flat space \cite{DiFrancesco:1997nk} to the de Sitter case. Starting from the de Sitter geometry and its conformal group, we derive Ward identities that constrain $n$-point correlators in conformally invariant states.

\subsection*{Conformal Killing vectors}
We look for the conformal Killing vectors $\epsilon^\mu$ that generate dS\textsubscript{d} conformal transformations. Since de Sitter space is conformally flat, we know that its conformal group is SO(2,$d$) as for $\mathbb{M}^{1,d-1}$. Leaving aside the isometry subgroup SO(1,$d$) $\subset$ SO(2,$d$), we focus on the remaining $d+1$ generators of `pure' conformal transformations. A coordinate transformation $x^{\mu}(x')=x'^{\mu}-\epsilon^{\mu}$ induced by a conformal Killing vector $\epsilon^\mu$  must satisfy $g'_{\mu\nu}(x)=e^{\omega(x)}g_{\mu\nu}(x)$, which infinitesimally reads
\begin{equation}
\mathcal{L}_{\epsilon}g_{\mu\nu}=\nabla_\mu \epsilon_\nu+\nabla_\nu \epsilon_\mu=-\omega(x)g_{\mu\nu}(x),
\end{equation} 
where $\mathcal{L}_\epsilon$ is the Lie derivative with respect to the Killing vector $\epsilon$. Assuming $\omega(x)\neq 0$ and contracting with the inverse metric $g^{\mu\nu}$ one obtains
\begin{equation}
\frac{2}{d}\ g_{\mu\nu}\nabla.\epsilon=\nabla_\mu \epsilon_\nu+\nabla_\nu \epsilon_\mu, \qquad \omega(x)=-\frac{2}{d}\nabla_{\mu}\epsilon^{\mu}.
\end{equation}
We find $d+1$ independent solutions to this equation parameterized by some constants $a, b, c^i$, which we express in conformal coordinates \eqref{conformalpatch}: 
\begin{align}
\label{conformalKillings}
(\epsilon^\eta,\epsilon^i)=a(\vec{x}^2+\eta^2,2\eta x^i)+b(1,0)+(\vec{c}.\vec{x},c^i\eta).
\end{align}
This transformation induces a Weyl rescaling of the metric that at first order reads
\begin{align}
\label{Omega}
\Omega(x)&\equiv e^{\omega(x)/2}\simeq\left(1+2a\eta-a\eta^{-1}(\eta^2+\vec{x}^2)-b\eta^{-1}-\vec{c}.\vec{x}\ \eta^{-1}\right).
\end{align}
The conformal Killing vectors \eqref{conformalKillings}, together with those generating de Sitter isometries, form a representation of the conformal algebra $so(2,d)$.

\subsection*{Matter Ward identities} \label{subsection:Ward}
Fundamental to the description of de Sitter CFTs are \textit{primary fields}, which transform in irreducible representations of the conformal group SO(2,$d$). For simplicity, we focus here on primary fields $\mathcal{O}_\Delta(x)$ transforming trivially (scalars) under the de Sitter isometry group SO(1,$d$). These are simply characterized by their scaling dimension $\Delta$. Under a conformal transformation $x^{\mu}(x')$ inducing a Weyl rescaling of the metric,
\begin{equation}
g'_{\mu\nu}(x)=\Omega(x)^2g_{\mu\nu}(x),
\end{equation}
primary fields transform as 
\begin{equation}
\mathcal{O}_{\Delta}'(x')=\Omega(x)^{-\Delta} \mathcal{O}_{\Delta}(x).
\end{equation}  
From this and \eqref{Omega} one can compute the first order functional variation of a scalar primary field in de Sitter space:
\begin{subequations}
\begin{align}
\delta \mathcal{O}_\Delta(x)&\equiv\mathcal{O}_\Delta'(x)-\mathcal{O}_\Delta(x) =\Omega(x-\epsilon)^{-\Delta}\mathcal{O}_\Delta(x-\epsilon)-\mathcal{O}_\Delta(x)\\
&=-\Delta\left(2a\eta-a\eta^{-1}(\eta^2+x^2)-b\eta^{-1}-c.x\eta^{-1}\right)\hat{\phi}-\mathcal{L}_\epsilon \mathcal{O}_\Delta(x),
\end{align}
\end{subequations}
which we decompose as
\begin{align}
\label{variation1}
\delta \mathcal{O}_\Delta(x)&=-a\left((\eta^2+x^2)(\partial_\eta-\frac{\Delta}{\eta})+2\eta\left(x^i\partial_i+\Delta\right)\right)\mathcal{O}_\Delta(x),\\
\label{variation2}
\delta \mathcal{O}_\Delta(x)&=-b\left(\partial_\eta-\frac{\Delta}{\eta}\right)\mathcal{O}_\Delta(x),\\
\label{variation3}
\delta \mathcal{O}_\Delta(x)&=-c^i\left(x_i\left(\partial_\eta-\frac{\Delta}{\eta}\right)+\eta\partial_i\right)\mathcal{O}_\Delta(x).
\end{align}
If the theory is conformal, one can insert \eqref{variation1}-\eqref{variation3} into the Schwinger-Dyson equations satisfied by $n$-point correlators of primary fields in conformally invariant states:
\begin{equation}
\label{Schwinger-Dyson}
\sum_{i=1}^{n} \langle \mathcal{O}_{\Delta_1}(n_1)...\delta \mathcal{O}_{\Delta_i}(n_i)...\mathcal{O}_{\Delta_n}(n_n)\rangle=0.
\end{equation}
This leads to a set of $d+1$ differential equations satisfied by such $n$-point correlators, the \textit{matter conformal Ward identities}. Together with SO(1,$d$) invariance, these differential equations completely fix the dependence of two- and three-point functions. For the two-point correlator we have:
\begin{align}
\label{Twopointconformal}
\langle \mathcal{O}_{\Delta_1}(n_1) \mathcal{O}_{\Delta_2}(n_2) \rangle=C_{\Delta_1} \delta_{\Delta_1,\Delta_2}\left(\frac{2}{1-n_1.n_2}\right)^{\Delta_1},
\end{align}
with $C_{\Delta_1}$ some normalization constant. We stress that conformal symmetry forbids a term of the form 
\begin{equation}
\left(\frac{2}{1+n_1.n_2}\right)^{\Delta},
\end{equation}
that could be obtained from \eqref{Twopointconformal} by replacing $n_1$ by its antipodal point $n_1^A=-n_1$. Similarly for three-point correlators conformal invariance imposes
\begin{align}
\nonumber
&\langle \mathcal{O}_{\Delta_1}(n_1) \mathcal{O}_{\Delta_2} (n_2) \mathcal{O}_{\Delta_3} (n_3) \rangle\\
&=C_{\Delta_1,\Delta_2,\Delta_3}\left(\frac{2}{1-n_1.n_2}\right)^{\frac{\Delta_1+\Delta_2-\Delta_3}{2}}\left(\frac{2}{1-n_1.n_3}\right)^{\frac{\Delta_1-\Delta_2+\Delta_3}{2}}\left(\frac{2}{1-n_2.n_3}\right)^{\frac{-\Delta_1+\Delta_2+\Delta_3}{2}}.
\end{align}
The interaction coefficient $C_{\Delta_1,\Delta_2,\Delta_3}$ depends on the theory under consideration. We should point out that a symmetry of the classical theory can be altered at the quantum level by \textit{anomalies}. However, for matter fields (by contrast to stress-energy) correlators, a conformal anomaly would modify the above functional forms at coincident points and for integer $\nu\equiv \Delta-\frac{d}{2}$ only \cite{Skenderis:2002wp}. In this paper we omit these subtleties.

\section{dS conformal group from AdS isometries} \label{section:conformalgroup}
Let's now show that AdS Killing vectors generate de Sitter conformal transformations on the conformal boundary $\mathcal{B}_d$. Bulk isometries are most easily described in embedding space $\mathbb{M}^{2,d}$, as they act as pseudorotations. Killing vectors generating these pseudorotations simply are
\begin{equation}
\label{Killing}
\epsilon_M=\omega_{MN}X^N, \qquad \omega_{MN}=-\omega_{NM}.
\end{equation}   
Writing these in de Sitter foliation \eqref{dS_slicing} and separating contributions from independent components of $\omega_{MN}$, \eqref{Killing} reduces to
\begin{align}
\label{Killing1}
\left(\epsilon^r,\epsilon^\alpha\right)&=(0,\omega^\alpha_{\ \ \beta}\ n^\beta),\\
\label{Killing2}
\left(\epsilon^r,\epsilon^\alpha\right)&=(\omega_{\beta d+1}\ n^\beta,\coth r\ \omega^\alpha_{\ \ d+1}).
\end{align}
One can readily see that \eqref{Killing1} generate isometries on dS\textsubscript{d} slices, irrespective of the value of $r$. This is completely similar to the usual Poincaré slicing of AdS in which Lorentz transformations are realized on any slice, not only on the flat conformal boundary. In order to see the emergence of de Sitter conformal Killing vectors as one approaches the boundary $r\to \infty$, we use conformal coordinates \eqref{conformalpatch} on constant $r$ slices. The components of \eqref{Killing2} in these coordinates further reduce to
\begin{align}
\left(\epsilon^r,\epsilon^\eta,\epsilon^i\right)&=\omega_{0d+1}\left(\frac{1}{2\eta}(\eta^2-\vec{x}^2-1),-\coth r\ \frac{1}{2}(\eta^2+\vec{x}^2+1), -\coth r\ \eta\ x^i\right),\\
\left(\epsilon^r,\epsilon^\eta,\epsilon^i\right)&=\omega_{dd+1}\left(\frac{1}{2\eta}(\eta^2-\vec{x}^2+1),-\coth r\ \frac{1}{2}(\eta^2+\vec{x}^2-1), -\coth r\ \eta\ x^i\right),\\
\left(\epsilon^r,\epsilon^\eta,\epsilon^i\right)&=\frac{1}{\eta}\left(\omega_{jd+1}x^j, \coth r\ \omega_{jd+1}x^j\ \eta, \coth r\ \omega^i_{\ d+1}\ \eta^2\right).
\end{align}
Taking independent combinations of the first two vectors and renaming the components of $\omega_{\alpha d+1}$, we obtain
\begin{align}
\label{bulkKilling1}
\left(\epsilon^r,\epsilon^\eta,\epsilon^i\right)&=a\left(\eta^{-1}(\vec{x}^2-\eta^2),\coth r\  (\vec{x}^2+\eta^2),2\coth r\ \eta\ x^i\right),\\
\label{bulkKilling2}
\left(\epsilon^r,\epsilon^\eta,\epsilon^i\right)&=b\left(\eta^{-1},\coth r\ ,0\right),\\
\label{bulkKilling3}
\left(\epsilon^r,\epsilon^\eta,\epsilon^i\right)&=\left(\eta^{-1}\ \vec{c}.\vec{x},\coth r\ \vec{c}.\vec{x}, c^i \coth r\ \eta\right).
\end{align}
As $r\to \infty$ and upon comparison with \eqref{conformalKillings}, one sees that these Killing vectors generate conformal transformations on the boundary $\mathcal{B}_d$. Their action in the radial direction is non-vanishing, but this should be expected since CFTs on distinct conformally flat metrics should be holographically represented by distinct choices of AdS conformal boundaries in the bulk, or equivalently distinct sections of the light-cone in embedding space $\mathbb{M}^{2,d}$. The radial action generated by \eqref{bulkKilling1}-\eqref{bulkKilling3} is therefore understood as moving the conformal boundary accordingly from one light-cone section to another. 

\section{Orthogonality of Legendre functions} \label{Appendix:Legendre}
We present new orthogonality relations for associated Legendre functions $P^{i\mu}_{\nu}$ and $Q^{i\mu}_\nu$ with lower index $\nu \neq-\frac{1}{2}$. The method presented here follows closely that of reference \cite{Bielsky}. We will make use of the following identities \cite{Bielsky,Magnus}: 
\begin{equation}
\label{Legendre1}
\left[ \partial_z\left((1-z^2)\partial_z\right)+\nu(\nu+1)+\frac{\mu^2}{1-z^2} \right] P_{\nu}^{i\mu}(z)=0,
\end{equation}
\begin{align}
\label{Legendre2}
(1-z^2)\partial_z P_{\nu}^{i\mu}(z)&=(\nu+1)zP_{\nu}^{i\mu}(z)-(\nu-i\mu+1)P_{\nu+1}^{i\mu}(z),\\
\label{Legendre3}
\lim\limits_{z\to \infty} P^{i\mu}_\nu(z)&=\frac{2^\nu\Gamma\left[\nu+\frac{1}{2}\right]}{\Gamma\left[\nu-i\mu+1\right]}z^\nu, \qquad \nu>-1/2,\\
\label{Legendre4}
\lim\limits_{z\to 1} P^{i\mu}_\nu(z)&=\frac{2^{i\mu/2}}{\Gamma\left[1-i\mu\right]}(z-1)^{-i\mu/2},\\
\label{Legendre5}
\lim\limits_{z\to 1^+} (z-1)^{-i(\mu+\mu')/2}&=i\pi (\mu+\mu')\delta(\mu+\mu').
\end{align}
Let's first restrict to $\nu >-\frac{1}{2}$. Using \eqref{Legendre1}-\eqref{Legendre2} it is shown that
\begin{align}
\nonumber
&(\mu'^2-\mu^2)\int_1^\infty \frac{dz}{1-z^2}P_{\nu}^{i\mu}(z)P_{\nu}^{i\mu'}(z)\\
\label{Ortho1}
&=\left[(\nu-i\mu'+1)P_{\nu}^{i\mu}(z)P_{\nu+1}^{i\mu'}(z)-(\nu-i\mu+1)P_{\nu}^{i\mu'}(z)P_{\nu+1}^{i\mu}(z)\right]_1^\infty.
\end{align}
Looking first at the limit $z\to \infty$ and using \eqref{Legendre3}, the first term gives
\begin{align}
\lim\limits_{z\to \infty}\ \left[(\nu-i\mu'+1)P_{\nu}^{i\mu}(z)P_{\nu+1}^{i\mu'}(z)-(\nu-i\mu+1)P_{\nu}^{i\mu'}(z)P_{\nu+1}^{i\mu}(z)\right]=0.
\end{align}
Using \eqref{Legendre4}-\eqref{Legendre5}, the second term of \eqref{Ortho1} in the limit $z\to 1$ gives
\begin{align}
\nonumber
&\lim\limits_{z\to 1}\ \left[(\nu-i\mu'+1)P_{\nu}^{i\mu}(z)P_{\nu+1}^{i\mu'}(z)-(\nu-i\mu+1)P_{\nu}^{i\mu'}(z)P_{\nu+1}^{i\mu}(z)\right]\\
&=\frac{\pi(\mu'^2-\mu^2) \delta(\mu+\mu')}{\Gamma\left[1-i\mu\right]\Gamma\left[1+i\mu\right]}.
\end{align}
We conclude for the orthogonality of associated Legendre functions of the first kind:
\begin{equation}
\int_1^\infty \frac{dz}{1-z^2}\ P_{\nu}^{i\mu}(z) P_{\nu}^{i\mu'}(z)=-\frac{\pi\delta(\mu+\mu')}{\Gamma\left[1-i\mu\right]\Gamma\left[1+i\mu\right]}=-\frac{\sinh \mu\pi}{\mu}\ \delta(\mu+\mu').
\end{equation}
Although we restricted to $\nu>-\frac{1}{2}$ for the computation, since $P^{i\mu}_{-\nu-1}(z)=P^{i\mu}_{\nu}(z)$ this orthogonality relation is actually valid for any real $\nu \neq -\frac{1}{2}$. From this and the relation between Legendre functions,
\begin{equation}
Q_{\nu}^{i\mu}(z)=-i \frac{\pi e^{-\mu\pi}}{2 \sinh \mu\pi}\left(P_\nu^{i\mu}(z)-\frac{\Gamma\left[\nu+i\mu+1\right]}{\Gamma\left[\nu-i\mu+1\right]}P^{-i\mu}_\nu(z)\right),
\end{equation}
one finds the orthogonality relation for associated Legendre functions of the second kind:
\begin{equation}
\label{Legendreortho}
\scalebox{1.1}{$\int_1^\infty \frac{dz}{1-z^2}\ Q_{\nu}^{i\mu}(z) Q_{\nu}^{i\mu'}(z)=-\frac{\pi^2}{2\mu\sinh\mu\pi}\left(\delta(\mu+\mu')+e^{-2\mu\pi} \frac{\Gamma\left[\nu+i\mu+1\right]}{\Gamma\left[\nu-i\mu+1\right]}\delta(\mu-\mu')\right).$}
\end{equation} 
In computing the orthonormality relation \eqref{orthonormality}, we have used \eqref{Legendreortho} with $\mu,\mu'>0$:
\begin{equation}
\int_1^\infty \frac{dz}{1-z^2}\ Q_{\nu}^{-i\mu}(z) Q_{\nu}^{i\mu'}(z)=-\frac{\pi^2\ \delta(\mu-\mu')}{2\mu \sinh\mu\pi}, \qquad \nu\neq -\frac{1}{2}.
\end{equation}

\section{Boundary-bulk propagators} \label{Appendix:Boundary-bulk}
In this appendix we explain how boundary-to-bulk propagators of a massive scalar field in AdS are obtained. Boundary-to-bulk propagators have to satisfy the following relation between bulk value of the field $\phi(X)$ and its imposed boundary value $\phi_{(d-\Delta)}(n)$:
\begin{equation}
\label{BB1}
\phi(X)=\int_{\mathcal{B}_d} d^dn'\ \sqrt{-g_{(0)}(n')}\ K(n',X)\ \phi_{(d-\Delta)}(n').
\end{equation}
Using Fefferman-Graham coordinates $X=(\rho,n)$ as defined in \eqref{Fefferman} and taking the limit $\rho \to 0$ we see that consistency requires
\begin{equation}
\lim\limits_{\rho \to 0} K(n',X) = \rho^{\frac{d-\Delta}{2}} \frac{\delta^d(n-n')}{\sqrt{-g_{(0)}(n)}}.
\end{equation}
One can expand equation \eqref{BB1} to higher order in $\rho$. At order $\rho^{\frac{\Delta}{2}}$ this leads to a useful formula for the field vev:
\begin{equation}
\phi_{(\Delta)}(n)=\int_{\partial M} d^dn'\ \sqrt{-g^{(0)}(n')}\ K_{(\Delta)}(n',n)\ \phi_{(d-\Delta)}(n').
\end{equation}
Boundary-to-bulk can be derived from bulk-to-bulk propagators by taking one insertion point to the boundary. To show this we make use of Green's second identity:
\begin{equation}
\int_{AdS} dV\ \left(\psi \square \phi -\phi \square \psi\right)= \int_{\mathcal{B}_d} d\Sigma^\mu\ \left(\psi \partial_\mu \phi-\phi \partial_\mu \psi\right).
\end{equation} 
Applied to $\psi(X)=K(n'',X)$ and $\phi(X)=G(X,X')$, the left-handside is
\begin{align}
\nonumber
\int_{AdS}& d^{d+1}X\ \sqrt{-g} \left(K(n'',X)\left(\square-m^2 \right)G(X,X')-G(X,X')\left(\square-m^2\right)K(n'',X)\right)\\
&=\int_M d^{d+1}X\ K(n'',X) \delta^{d+1}(X-X')=K(n'',X'),
\end{align}
while the right-handside is, assuming $G$ to have leading normalizable behavior at the boundary,
\begin{subequations}
\begin{align}
\nonumber
-2\int_{\mathcal{B}_d}&d^dn\ \sqrt{-g_{(0)}(n)}\ \rho^{1-d/2} \left(K(n'',X)\partial_\rho G(X,X')-G(X,X')\partial_\rho K(n'',X)\right)\\
&=(d-2\Delta) \int_{\mathcal{B}_d} d^dn\ \sqrt{-g_{(0)}(n)}\ \rho^{-d/2} K(n'',X) G(X,X')\\
&=(d-2\Delta) \lim\limits_{\rho'' \to 0} (\rho'')^{-\frac{\Delta}{2}}\ G(X'',X').
\end{align} 
\end{subequations}
From this we deduce 
\begin{align}
\label{boundary-bulk}
K^{AdS}_{\alpha_\mu}(n',X)&=-2\nu\lim\limits_{\rho'\rightarrow 0} \left(\rho'\right)^{-\frac{\Delta}{2}} G_{\alpha_\mu}^{AdS}(X',X).
\end{align}

\subsubsection*{Consistency Check}
Obtaining the expression of $K^{AdS}_{\alpha_\mu}(n',X)$ in closed form is in general difficult because of the $\mu$ integration in \eqref{modesum} that one should perform. As example and consistency check, we obtain a closed form expression for the boundary-to-bulk propagator in the conformal vacuum $|E\rangle$ ($\alpha_\mu=-\infty$) and odd dimension $d$, by two independent methods: the first one consists in explicitly performing the mode summation \eqref{modesum} in the limit of \eqref{boundary-bulk}; in the second one uses the well-known closed form of the unique propagator in EAdS and takes the limit \eqref{boundary-bulk}. These expressions have to agree upon analytic continuation from Lorentzian to Euclidean signature. From the radial function $\psi_\mu(z)$ given in \eqref{radialfunction1} or \eqref{radialfunction2} and the large argument limit of the Legendre function \cite{Magnus},
\begin{equation}
\lim\limits_{z \to \infty} Q_{\nu-\frac{1}{2}}^{i\mu}(z)=e^{-\mu\pi}\frac{\sqrt{\pi}\Gamma\left[\nu+i\mu+\frac{1}{2}\right]}{2^{\nu+\frac{1}{2}}\Gamma\left[\nu+1\right]}z^{-(\nu+\frac{1}{2})} \qquad \nu\neq -1,
\end{equation}
we have
\begin{subequations}
\begin{IEEEeqnarray}{l}
\nonumber	
\vspace{0.2cm}
K_E^{AdS}(P,X)\\
=-2\nu \lim\limits_{r' \to \infty} (2\cosh r')^{\Delta}\ G_E^{AdS}(X',X)\\
\scalebox{0.99}{$=-\frac{2^{\frac{d+3}{2}}\nu}{\pi^{3/2}\Gamma\left[\nu+1\right]} (\sinh r)^{\frac{1-d}{2}} \int_0^\infty d\mu\ \mu \sinh\mu\pi\ e^{\mu\pi}\ \Gamma\left[\nu-i\mu+\frac{1}{2}\right] Q_{\nu-\frac{1}{2}}^{i\mu}(\cosh r)G_E^{dS}(n,n').$}
\IEEEeqnarraynumspace
\end{IEEEeqnarray}
\end{subequations}
We will make use of expression \eqref{TwopointGegenbauer} for the de Sitter propagator as well as the two following formulas \cite{Gradshteyn,Grosche:1987de}:
\begin{equation}
Q_{\nu-\frac{1}{2}}^{i\mu}(\cosh r)=\frac{e^{-\mu\pi}\Gamma\left[\nu+\frac{1}{2}\right]}{\Gamma\left[\nu-i\mu+\frac{1}{2}\right]} \int_0^\infty dt\ \frac{\cos \mu t}{\left(\cosh r+\sinh r\ \cosh t\right)^{\nu+\frac{1}{2}}},
\end{equation}
\begin{equation}
-i\mu\ C^{\frac{d-1}{2}}_{-\frac{d-1}{2}-i\mu}(\cosh \gamma) =\frac{2^{\frac{3-d}{2}}} {\Gamma\left[\frac{d-1}{2}\right]} \left(\frac{d}{d\cosh \gamma}\right)^{\frac{d-1}{2}} \cos \mu \gamma, \qquad \gamma>0.
\end{equation}
First we perform the $\mu$ integration,
\begin{IEEEeqnarray}{l}
\int_0^\infty d\mu\ \cos \mu t\ \cos \mu\gamma=\frac{1}{2}\left[\frac{\sin \mu(t+\gamma)}{t+\gamma}+\frac{\sin \mu(t-\gamma)}{t-\gamma}\right]_0^\infty=\frac{\pi}{2}\left(\delta(t+\gamma)+\delta(t-\gamma)\right).
\IEEEeqnarraynumspace
\end{IEEEeqnarray}
Thanks to the delta distributions, integration over the $t$ variable is trivial and we simply have to compute 
\begin{IEEEeqnarray}{l}
\left(\frac{d}{d\cosh\gamma}\right)^{\frac{d-1}{2}} \frac{1}{\left(\cosh r+\sinh r\ \cosh \gamma\right)^{\nu+\frac{1}{2}}}=\frac{(-1)^{\frac{d-1}{2}}\Gamma\left[\Delta\right]}{\Gamma\left[\nu+\frac{1}{2}\right]} \frac{(\sinh r)^{\frac{d-1}{2}}}{\left(\cosh r-\sinh r\ n.n'\right)^{\Delta}}.
\IEEEeqnarraynumspace
\end{IEEEeqnarray}
\newline
Collecting the results, we end up with
\begin{align}
\label{BBeuclidean}
K_E^{AdS}(P,X)&=-\frac{\Gamma\left[\Delta\right]}{\pi^{d/2}\Gamma\left[\nu\right]} \frac{1}{\left(\cosh r-\sinh r\ n.n'\right)^{\Delta}}=-\frac{\Gamma\left[\Delta\right]}{\pi^{d/2}\Gamma\left[\nu\right]} \frac{1}{\left(-P.X\right)^{\Delta}},
\end{align}
where 
\begin{equation}
\label{XP}
P_1.X_2=\eta_{MN}P^MX^N=l \cosh r_2 \left(\tanh r_2\ n_1.n_2-1\right).
\end{equation}
The same expression can be derived from the well-know closed form of the Euclidean bulk-to-bulk propagator in AdS\textsubscript{d+1},
\begin{align}
G_E^{AdS}(X,X')=\frac{\Gamma\left[\Delta\right]}{2^{\nu+1}(2\pi)^{d/2}\Gamma\left[\nu+1\right]}\ \xi^{\Delta}\ {}_2F_1\left[\frac{\Delta}{2},\frac{\Delta+1}{2},1-\frac{d}{2}+\Delta;\xi^2\right],
\end{align}
where $\xi\equiv -X.X'$. Taking the limit \eqref{boundary-bulk} leads to \eqref{BBeuclidean}.

\section{Euclidean three-point correlators} \label{Appendix:ThreePoint}
In this appendix we illustrate the computation of higher $n$-point correlators with the three-point correlator of primary operators in the conformal vacuum. For this one should add a cubic vertex interaction of the form $\mathcal{L}_{int}^{AdS}=\phi_{\Delta_1}\phi_{\Delta_2}\phi_{\Delta_3}$ in the bulk of AdS. This correlator is computed from the Feynman-Witten diagram shown in Figure \ref{fig:Witten} where the three Euclidean boundary-to-bulk propagators $K^{AdS}_E$ associated to each bulk field meet at any bulk point (which is integrated over). See \cite{Freedman:1998tz} for the flat boundary computation. For later use we define the invariant distance between two boundary points $P_1, P_2 \in \mathcal{B}_d$:
\begin{equation}
P_{12}\equiv (P_1-P_2)^2=-2P_1.P_2=2l^2(1-n_1.n_2).
\end{equation}
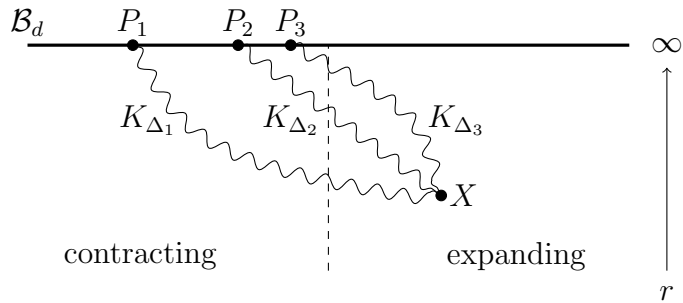
\begin{figure}
	\centering
	\begin{tikzpicture}
	\draw[->] (8.5,2) -- (8.5,4.7);
	\draw (8.5,1.7) node {$r$};
	\draw (8.5,5) node {$\infty$}; 
	\draw (0,5.3) node {$\mathcal{B}_d$};
	\draw[very thick] (0,5) -- (8,5);
	\filldraw [black] (1.4,5) circle (2pt) (2.8,5) circle (2pt) (3.5,5) circle (2pt) (5.5,3) circle (2pt);
	\draw (1.4,5.3) node {$P_1$} (2.8,5.3) node {$P_2$} (3.5,5.3) node {$P_3$} (5.8,3) node {$X$};
	\draw[snake it] (3.5,5) .. controls (5.5,4) and (5,4) .. (5.5,3);
	\draw (5.7,4) node {$K_{\Delta_3}$};
	\draw[snake it] (2.8,5) -- (5.5,3);
	\draw (3.5,4) node {$K_{\Delta_2}$};
	\draw[snake it] (1.4,5) .. controls (2,4) and (2,3.5) .. (5.5,3);
	\draw (1.6,4) node {$K_{\Delta_1}$};
	\draw[dashed] (4,2) -- (4,5);
	\draw (1.5,2.2) node {contracting} (6.5,2.2) node {expanding};
	\end{tikzpicture}
	\caption{Feynman-Witten diagram representing the holographic computation of three-point correlators in the boundary CFT. The three insertion points $P_1, P_2, P_3 \in \mathcal{B}_d$ are in this case located in the contracting patch of the de Sitter boundary. The bulk interaction point $X \in AdS_{d+1}$ is integrated in the bulk over global de Sitter slices (both expanding and contracting patches).}
	\label{fig:Witten}
\end{figure}
The relevant quantity to compute is
\begin{subequations}
\begin{align}
A(P_1,P_2,P_3)&=\int_{AdS} dX \prod_{i=1,2,3} K_{\Delta_i}(P_i,X)\\
&=-\frac{1}{\pi^{3d/2}}\int_{AdS} dX \prod_{i=1,2,3}\frac{\Gamma\left[\Delta_i\right]}{\Gamma\left[\nu_i\right]} \left(\frac{1}{-X.P_i}\right)^{\Delta_i}.
\end{align}
\end{subequations}
The method we will use is completely parallel to the one presented in \cite{Penedones:2016voo} and makes clever use of AdS isometries. Using Schwinger parametrization
\begin{equation}
\left(\frac{1}{-X.P}\right)^{\Delta}=\frac{1}{\Gamma\left[\Delta\right]}\int_{0}^{\infty} ds\ s^{\Delta-1} e^{sX.P},
\end{equation}
we then have to perform the $AdS$ integral
\begin{align}
\int_{AdS} dX\ e^{X.Q},
\end{align}
with $Q\equiv s_1P_1+s_2P_2+s_3P_3$. Since $Q^2=-\left(s_1s_2P_{12}+s_1s_3P_{13}+s_2s_3P_{23}\right)<0$ for spacelike separated points $P_1,P_2,P_3 \in \mathcal{B}_d$, then $Q$ is a timelike vector.\footnote{If $Q$ is spacelike we can align it to $X^d$ and the derivation goes through.} We express the bulk metric in de Sitter foliation \eqref{dS_slicing} and contracting coordinates \eqref{conformalpatch} on the slices. Using SO(2,$d$) symmetry we align $Q$ to $X^0$, $Q=-|Q|e^{0}$, which implies
\begin{subequations}
\begin{align}
\int_{AdS} dX\ e^{X.Q}&=\int d(\cosh r) \int_{0}^{\infty} \frac{d\eta}{\eta} \int_{\mathbb{R}^{d-1}} d^{d-1}\vec{x}\ \left(\frac{\sinh r}{\eta}\right)^{d-1} e^{\frac{\sinh r}{2\eta}(\eta^2-x^2-1)|Q|}\\
&=\pi^{\frac{d-1}{2}}\int d(\cosh r)\int_{0}^{\infty} \frac{d\eta}{\eta}\ \left(\frac{2\sinh r}{\eta |Q|}\right)^{\frac{d-1}{2}} e^{\frac{\sinh r}{2\eta}(\eta^2-1)|Q|}.
\end{align}
\end{subequations}
Performing the variable changes $z=\cosh r$ and $\bar{\eta}=\frac{\eta|Q|}{2\sinh r}$, we get\footnote{From the definition $z=\cosh r$ it would seem like the $z$ variable should run from $1$ to $\infty$ only. However this coordinate system only covers the part of AdS with $X^{d+1}>1$. Using an anti-de Sitter foliation for the region $0<X^{d+1}<1$, one can check that the exact same integral is obtained with integration range $z \in \left[0,1\right]$.}
\begin{subequations}
\begin{align}
\int_{AdS} dX\ e^{X.Q}&=\pi^{\frac{d-1}{2}}\int_{0}^{\infty} \frac{d\bar{\eta}}{\bar{\eta}}\ \left(\frac{1}{\bar{\eta}}\right)^{\frac{d-1}{2}}e^{Q^2/4\bar{\eta}}\ e^{-\bar{\eta}} \int_{0}^{\infty}dz\ e^{z^2\bar{\eta}}\\
&= \frac{i\pi^{\frac{d}{2}}}{2}\int_{0}^{\infty} \frac{d\bar{\eta}}{\bar{\eta}}\ \bar{\eta}^{-\frac{d}{2}}\ e^{Q^2/4\bar{\eta}}\ e^{-\bar{\eta}}.
\end{align}
\end{subequations}
In order to factorize the integrals over $s_1, s_2, s_3$, we apply the variable changes $s_i=\frac{\sqrt{\bar{\eta}t_1t_2t_3}}{t_i}$:
\begin{subequations}
\begin{align}
A(P_1,P_2,P_3)&=-\frac{i}{4\pi^d}\frac{\Gamma\left[\frac{\Delta_1+\Delta_2+\Delta_3-d}{2}\right]}{\Gamma\left[\nu_1\right]\Gamma\left[\nu_2\right]\Gamma\left[\nu_3\right]}\int_{0}^{\infty} \frac{dt_1}{t_1}\ t_1^{\frac{-\Delta_1+\Delta_2+\Delta_3}{2}}e^{-t_1P_{23}/4}\int_{0}^{\infty} \frac{dt_2}{t_2}\ ...\\
\label{threepoint}
&=\frac{a}{2}\ \frac{2^{\Delta_1+\Delta_2+\Delta_3}}{P_{12}^{\frac{\Delta_1+\Delta_2-\Delta_3}{2}}P_{13}^{\frac{\Delta_1-\Delta_2+\Delta_3}{2}}P_{23}^{\frac{-\Delta_1+\Delta_2+\Delta_3}{2}}},
\end{align}
\end{subequations}
\begin{equation}
a\equiv -\frac{i}{2 \pi^d}\frac{\Gamma\left[\frac{\Delta_1+\Delta_2+\Delta_3-d}{2}\right]\Gamma\left[\frac{\Delta_1+\Delta_2-\Delta_3}{2}\right]\Gamma\left[\frac{\Delta_1-\Delta_2+\Delta_3}{2}\right]\Gamma\left[\frac{-\Delta_1+\Delta_2+\Delta_3}{2}\right]}{\Gamma\left[\nu_1\right]\Gamma\left[\nu_2\right]\Gamma\left[\nu_3\right]}.
\end{equation}
We still need to multiply the final result by a factor of 2 since the computation missed all contributions from expanding patches of bulk de Sitter slices. Indeed, even if all insertion points are located in the contracting patch of the boundary, points in expanding patches of dS slices located deeper in the bulk are still finite distance away. This can be seen from the scalar product $X.P$ given in \eqref{XP} where $P$ is a point in the contracting patch of the boundary $\mathcal{B}_d$ and $X$ is a point in the expanding patch of any bulk slice. Therefore such intermediate bulk points where the interaction takes place also contribute to the three-point function. See Figure \ref{fig:Witten}. Similarly to what has been discussed in Section \ref{subsection:conformalvacuum}, from \cite{Freedman:1998tz} one can check that \eqref{threepoint} is related to the flat space result by a conformal transformation (numerical factors included).

\section{De Sitter entanglement entropy} \label{Appendix:Entanglement}
One computes the entanglement entropy associated to some bipartition of the boundary theory from the AdS bulk geometry as the minimal surface area whose intersection with the boundary is precisely the bipartition surface. This so-called Ryu-Takayanagi proposal \cite{Ryu:2006bv} can be understood in a more general framework as bulk gravitational entropy \cite{Hawking:1998jf,Lewkowycz:2013nqa}. As an illustration, we derive the de Sitter entropy arising from tracing out degrees of freedom living behind one inertial  observer's horizon (cosmological horizon). This generalizes the computation of \cite{Hawking:2000da,Li:2011bt} valid for $d=2,4$ to generic dimension $d$. Note that this entanglement entropy is also identified with the boundary gravitational entropy whose value is given by the well-known Gibbons-Hawking formula \cite{Gibbons:1977mu}:
\begin{equation}
\label{Bekenstein}
S=\frac{\text{Horizon}_{\text{dS\textsubscript{d}}}}{4G_d},
\end{equation}
where $\text{Horizon}_{\text{dS\textsubscript{d}}}$ is the area of the cosmological horizon. As important ingredient we need to identify the effective boundary Newton constant $G_d$ from bulk quantities. Assuming the Fefferman-Graham gauge \eqref{Fefferman} and expressing bulk geometrical quantities in terms of the radial coordinate $r$ and the transverse metric $g_{ij}$, we can reduce the bulk Einstein-Hilbert action to
\begin{align}
S_{EH}=\frac{l^{2-d}}{16\pi G_{d+1}}\int_{0}^{r_c\to \infty} dr\ \sinh^{d-2} r \int d^dx\ \sqrt{-g}R+...,
\end{align}
such that we identify the effective Newton constant $G_d$ as
\begin{equation}
\frac{1}{G_d}=\frac{l^{2-d}}{G_{d+1}} \int_{0}^{r_c \to \infty} dr\ \sinh^{d-2} r.
\end{equation} 
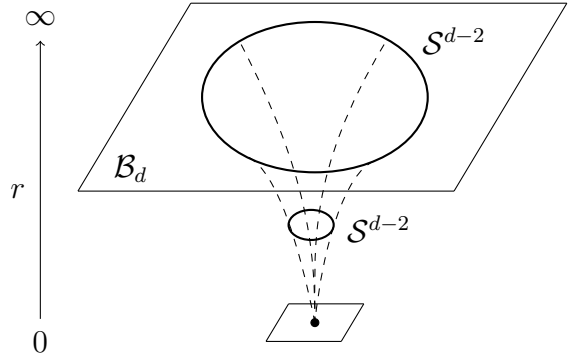
\begin{figure}
\centering
\begin{tikzpicture}[x={(1,0)},y={(0,1)},z={(0.3,0.5)}]
\draw[->] (-0.5,-1.7) -- (-0.5,2);
\draw (-0.5,-2) node {$0$};
\draw (-0.5,2.3) node {$\infty$};
\draw (-0.8,0) node {$r$}; 
\draw (0.7,0.3) node {$\mathcal{B}_d$};
\draw (0,0) -- (5,0);
\draw (0,0) -- (0,0,5);
\draw (5,0) -- (5,0,5);
\draw (0,0,5) -- (5,0,5);
\draw[thick] (2.4,0,2.5) ellipse (1.5 and 1); 
\draw (5,2) node {$\mathcal{S}^{d-2}$};
\draw[thick] (2.35,-1.7,2.5) ellipse (0.3 and 0.2);
\draw (4,-0.5) node {$\mathcal{S}^{d-2}$};
\filldraw [black] (2.4,-3,2.5) circle (1.5pt);
\draw (2.5,-2) -- (3.5,-2);
\draw (2.5,-2) -- (2.5,-2,1);
\draw (3.5,-2) -- (3.5,-2,1);
\draw (2.5,-2,1) -- (3.5,-2,1);
\draw[dashed] (2,0,0.9) .. controls (1.9,-1,2.5) and (2.2,-2,2.5) .. (2.4,-3,2.5);
\draw[dashed] (3.7,0,0.8) .. controls (2.8,-1,2.5) and (2.5,-2,2.5) .. (2.4,-3,2.5);
\draw[dashed] (2.9,0,3.9) .. controls (2.3,-1,2.5) and (2.4,-2,2.5) .. (2.4,-3,2.5);
\draw[dashed] (1,0,3.9) .. controls (2.3,-1,2.5) and (2.4,-2,2.5) .. (2.4,-3,2.5);
\end{tikzpicture}
\caption{The bulk Killing horizon in AdS\textsubscript{d+1} is the union of spheres $\mathcal{S}^{d-2}$ of radius $\sinh^2 r$. Its associated gravitational entropy $S=\frac{A}{4G_{d+1}}$ reproduces exactly the gravitational entropy of the cosmological horizon in the de Sitter boundary $\mathcal{B}_d$.} 
\label{fig:entropy}
\end{figure}
Next we turn to the computation of boundary gravitational entropy from bulk gravitational entropy. As exposed in \cite{Hawking:1998jf}, if Hamiltonian time evolution does not succeed in generating a flow through a foliation of spacetime by constant time hypersurfaces, then a breakdown of unitarity occurs in the quantum theory and associated entropy is generated. This happens in particular when Killing horizons left fixed (in spacetime) under time evolution are present. The entropy is then given by the quarter area formula $S=A/4G$, where $A$ is the Killing horizon area. Let's look at the coordinate system of one inertial observer called the static patch of de Sitter space, which is defined for $u\equiv\left(n_in^i\right)^{1/2} \in \left[0,1\right]$ and $n^d>0$:
\begin{align}
n^0&=\left(1-u^2\right)^{1/2}\sinh t,\\
n^d&=\left(1-u^2\right)^{1/2}\cosh t,\\
n^i&=u \bar{n}^i, \quad i=1,...,d-1,
\end{align}
\begin{equation}
\label{staticpatch}
dn^2=-\left(1-u^2\right)dt^2+\left(1-u^2\right)^{-1}du^2+u^2d\Omega^2_{d-2}.
\end{equation}
The inertial observer is located at $u=0$, and there is a Killing horizon at $u=1$ where the Killing vector $\partial_t$ has vanishing norm. The observer horizon is left fixed under time evolution and from the quarter area formula one concludes that the associated gravitational entropy is indeed given by \eqref{Bekenstein}. What we want is to derive the same expression in terms of AdS bulk gravitational entropy. For this we need to find a spacelike hypersurface in AdS\textsubscript{d+1} which is left invariant under the action of $\partial_t$. Thanks to the foliation \eqref{dS_slicing}, the answer is completely obvious: it is the union of  spheres $\mathcal{S}^{d-2}$ of radius $\sinh^2 r$ that satisfy 
\begin{equation}
\label{union}
X^0=X^d=0, \qquad \sum_{i=1}^{d-1}(X^i)^2=\sinh^2 r, \qquad X^{d+1}=\cosh r.
\end{equation}
See Figure \ref{fig:entropy}. Extending the static patch \eqref{staticpatch} throughout the bulk of AdS, this hypersurface is the union of cosmological horizons on each de Sitter slice. It has been proven in \cite{Lewkowycz:2013nqa} that $U(1)$-invariant hypersurfaces must have minimal surface, which makes direct connection with the Ryu-Takayanagi proposal. For $d=4$, it has indeed been checked that \eqref{union} satisfies the minimal area criterion \cite{Li:2011bt}. Parameterizing this hypersurface with angular variables $\left(r,\theta_1,...,\theta_{d-2}\right)$ and writing down the components of its induced metric $\gamma$,
\begin{align}
\gamma_{rr}&=\eta_{MN}\frac{\partial X^M}{\partial r}\frac{\partial X^N}{\partial r}=1, \qquad \gamma_{\theta_1\theta_2}=\sinh^2 r\ \sum_{i=1}^{d-1} \frac{\partial n^i}{\partial \theta_1}\frac{\partial n^i}{\partial \theta_2},\\
d^{d-1}\sigma&=\sinh^{d-2} r\ d\Omega_{d-2},
\end{align}
we conclude that its area is
\begin{subequations}
\begin{align}
A&=\int d^{d-1}\sigma=\omega_{d-2} \int_0^{r_c}dr\ \sinh^{d-2} r\\
&=\text{Horizon}_{\text{dS\textsubscript{d}}}\ l^{2-d}\int_0^{r_c}dr\ \sinh^{d-2} r=\text{Horizon}_{\text{dS\textsubscript{d}}} \frac{G_{d+1}}{G_d},
\end{align}
\end{subequations}
where $\omega_{d-2}$ is the volume of the unit sphere $\mathcal{S}^{d-2}_{unit}$. We conclude that the bulk gravitational entropy computed from the quarter area formula exactly reproduces the value of the de Sitter horizon entropy:
\begin{align}
S=\frac{A}{4G_{d+1}}=\frac{\text{Horizon}_{\text{dS\textsubscript{d}}}}{4G_d}.
\end{align}

\end{document}